\documentclass[review,12pt]{elsarticle}

\usepackage{amssymb}
\usepackage{amsthm}
\usepackage{amsmath}

\usepackage{mathrsfs}
\usepackage{graphicx}
\usepackage{epstopdf}
\usepackage{float}
\usepackage{caption}
\usepackage{subcaption}
\usepackage{bm}
\usepackage{bbm}
\usepackage{mathrsfs}
\usepackage{hyperref}
\usepackage{cleveref}
\usepackage{soul}
\usepackage{accents}
\usepackage{graphicx}
\usepackage{siunitx} % To define \mu as non-italic
\usepackage{xcolor}
\usepackage[margin=2cm]{geometry}% TO PLAY WITH THE MARGINS
\usepackage{courier} % To write in Courier font (to write code basically)
\usepackage{listings} % To include code
\usepackage{tabu} % To increase the vertical space between cells in tables
\usepackage{longtable}
\usepackage{framed} % To put a box to the nomenclature
\usepackage{changepage} % To use the adjustwidth feature that allows pushin wide tables to the left margin
\biboptions{sort&compress} %To make [1,2,3] be [1-3]
\usepackage{url} % just for breaking urls

\usepackage{graphicx}
\usepackage{array}
\usepackage{setspace}

\usepackage{nomencl}

\hypersetup{
    colorlinks=true,
    linkcolor=blue,
    filecolor=magenta,      
    urlcolor=cyan,
    }

\journal{ }

\makeatletter
\def\@author#1{\g@addto@macro\elsauthors{\normalsize%
    \def\baselinestretch{1}%
    \upshape\authorsep#1\unskip\textsuperscript{%
      \ifx\@fnmark\@empty\else\unskip\sep\@fnmark\let\sep=,\fi
      \ifx\@corref\@empty\else\unskip\sep\@corref\let\sep=,\fi
      }%
    \def\authorsep{\unskip,\space}%
    \global\let\@fnmark\@empty
    \global\let\@corref\@empty  %% Added
    \global\let\sep\@empty}%
    \@eadauthor={#1}
}
\makeatother

\begin{document}

\begin{frontmatter}



\title{A COMSOL framework for predicting hydrogen embrittlement - Part I: coupled hydrogen transport}


\author[UBU]{Andrés Díaz\corref{cor1}}
\ead{adportugal@ubu.es}

\author[UBU]{Jesús Manuel Alegre}
\author[UBU]{Isidoro Iván Cuesta}

\author{Emilio Mart\'{\i}nez-Pa\~neda\corref{cor1}\fnref{Ox}}
\ead{emilio.martinez-paneda@eng.ox.ac.uk}

\address[UBU]{University of Burgos, Escuela Politécnica Superior, 09006 Burgos, Spain}
\address[Ox]{Department of Engineering Science, University of Oxford, Oxford OX1 3PJ, UK}

\cortext[cor1]{Corresponding author.}

\begin{abstract}
Hydrogen threatens the structural integrity of metals and thus predicting hydrogen-material interactions is key to unlocking the role of hydrogen in the energy transition. Quantifying the interplay between material deformation and hydrogen diffusion ahead of cracks and other stress concentrators is key to the prediction and prevention of hydrogen-assisted failures. In this work, a generalised theoretical and computational framework is presented that for the first time encompasses: (i) stress-assisted diffusion, (ii) hydrogen trapping due to multiple trap types, rigorously accounting for the rate of creation of dislocation trap sites, (iii) hydrogen transport through dislocations, (iv) equilibrium (Oriani) and non-equilibrium (McNabb-Foster) trapping kinetics, (v) hydrogen-induced softening, and (vi) hydrogen uptake, considering the role of hydrostatic stresses and local electrochemistry. Particular emphasis is placed on the numerical implementation in COMSOL Multiphysics, releasing the relevant models and discussing stability, discretisation and solver details. Each of the elements of the framework is independently benchmarked against results from the literature and implications for the prediction of hydrogen-assisted fractures are discussed. The second part of this work (Part II) shows how these crack tip predictions can be combined with crack growth simulations.\\

\end{abstract}


\end{frontmatter}


\makenomenclature

% Nomenclature
% The following order should be used within this table: Latin characters should appear first, arranged a, A, b, B etc.; then Greek characters, similarly arranged; sub/superscripts, abbreviations, special functions etc. usually come as a separate final group.
% makeindex EFM.nlo -s nomencl.ist -o EFM.nls
\begin{framed}
\nomenclature{\(C\)}{Total hydrogen concentration}
\nomenclature{\(C_L\)}{Hydrogen concentration in lattice sites}

\nomenclature{\(C_T\)}{Hydrogen concentration in trapping sites}

\nomenclature{\(C_T^i\)}{Hydrogen concentration in different trapping sites. $i=m$: mobile dislocations; $i=d$: dislocations; $i=c$: carbides; $i=gb$: grain boundaries;}

\nomenclature{\(\mathbf{J}\)}{Hydrogen flux vector}
\nomenclature{\(\mathbf{J}_L\)}{Hydrogen flux vector through lattice sites}
\nomenclature{\(t\)}{Time}
\nomenclature{\(D_L\)}{Ideal diffusivity through lattice sites}
\nomenclature{\(R\)}{Constant of gases}
\nomenclature{\(T\)}{Temperature}
\nomenclature{\(\sigma_h\)}{Hydrostatic stress}
\nomenclature{\(\bar{V}_H\)}{Partial molar volume of hydrogen in the host metal}
\nomenclature{\(\theta_L\)}{Occupancy of lattice sites defined as $C_L/N_L$}
\nomenclature{\(\theta_T\)}{Occupancy of trapping sites defined as $C_T/N_T$}
\nomenclature{\(N_L\)}{Number of lattice sites per unit volume}
\nomenclature{\(N_T\)}{Number of trapping sites per unit volume (trap density)}
\nomenclature{\(K_T\)}{Equilibrium constant for trapping}
\nomenclature{\(E_B\)}{Binding energy for trapping}
\nomenclature{\(\varepsilon_p\)}{Equivalent plastic strain}
\nomenclature{\(D_{eff}\)}{Local effective diffusivity due to trapping effects}
\nomenclature{\(\bar{D}\)}{Operational non-dimensional diffusivity defined as $D_L/D_{eff}$}
\nomenclature{\(\kappa\)}{Capture constant for kinetic trapping}
\nomenclature{\(\lambda\)}{Release constant for kinetic trapping}
\nomenclature{\(\mathbf{J}^d\)}{Hydrogen flux vector due to transport by dislocations}
\nomenclature{\(\mathbf{v}^d\)}{Dislocation velocity vector}

\nomenclature{\(\rho^m\)}{Density of mobile dislocations}
\nomenclature{\(\gamma\)}{Geometric parameter correlating dislocation and trap densities}
\nomenclature{\(a\)}{Lattice parameter}
\nomenclature{\(b_v\)}{Burgers vector}
\nomenclature{\(\mathbf{n}\)}{Unitary vector}
\nomenclature{\(\mathcal{B}\)}{Surface of volume where boundary conditions are applied}
\nomenclature{\(C_{env}\)}{Concentration of lattice hydrogen in equilibrium with the environment at a given pressure}
\nomenclature{\(C_L^0\)}{Initial hydrogen concentration in lattice sites}
\nomenclature{\(K\)}{Solubility of lattice hydrogen}
\nomenclature{\(p_{H_2}\)}{Gaseous hydrogen pressure}
\nomenclature{\(\mu_L\)}{Chemical potential of hydrogen in lattice sites}
\nomenclature{\(\mu_{H_2}\)}{Chemical potential of gaseous hydrogen}
\nomenclature{\(\mu_L^0\)}{Reference chemical potential of hydrogen in lattice sites}
\nomenclature{\(\mu^0_{H_2}\)}{Reference chemical potential of gaseous hydrogen}
\nomenclature{\(p^0\)}{Reference pressure}
\nomenclature{\(f_{H_2}\)}{Fugacity of gaseous hydrogen}
\nomenclature{\(k_{abs}\)}{Absorption constant for electrochemical hydrogen uptake}
\nomenclature{\(k_{abs}^*\)}{Absorption constant in velocity units for electrochemical hydrogen uptake}
\nomenclature{\(k_{des}\)}{Desorption constant for electrochemical hydrogen uptake}
\nomenclature{\(k_c\)}{Charging constant for electrochemical hydrogen uptake}
\nomenclature{\(k_{r,elec}\)}{Electrochemical recombination constant for electrochemical hydrogen uptake}
\nomenclature{\(k_{r,chem}\)}{Chemical recombination constant for electrochemical hydrogen uptake}
\nomenclature{\(\theta_{ad}\)}{Coverage of adsorbed hydrogen for electrochemical uptake}

\nomenclature{\(\textbf{x}\)}{Coordinates vector in a spatial frame}
\nomenclature{\(\textbf{X}\)}{Coordinates vector in a material frame}
\nomenclature{\(\nabla\)}{Gradient operator in a spatial frame}
\nomenclature{\(\nabla_\textbf{X}\)}{Gradient operator in a material or reference frame}
\nomenclature{\(\Gamma_{ext}\)}{External flux term in the \textit{Transport in Solids} module}

\nomenclature{\(\sigma_y\)}{Yield stress}
\nomenclature{\(\sigma_{y0}\)}{Initial yield stress before hardening}
\nomenclature{\(f_s(c)\)}{Hydrogen-induced softening law}
\nomenclature{\(h(\varepsilon_p)\)}{Plastic strain hardening function}
\nomenclature{\(c\)}{Local hydrogen concentration in H atoms per metal atom}
\nomenclature{\(\zeta\)}{Coefficient for hydrogen-induced softening}
\nomenclature{\(\xi\)}{Coefficient for hydrogen-induced softening ($1-\zeta$)}

\nomenclature{\(E\)}{Young's modulus}
\nomenclature{\(N\)}{Strain hardening exponent}
\nomenclature{\(\mathbf{u}\)}{Displacement vector, with components $u_x$ and $u_y$ (in 2D)}
\nomenclature{\(R_T\)}{Reaction rate term in the governing equation of \textit{Transport of Diluted Species}}
\nomenclature{\(\mathbf{v}\)}{Convection velocity field in the governing equation of \texttt{Transport of Diluted Species}}
\nomenclature{\(J_{in}\)}{Hydrogen flux value for flux boundary conditions}

\nomenclature{\(d_a\)}{Damping term in the \textit{Stabilized convection-diffusion equation}}
\nomenclature{\(c_d\)}{Diffusion term in the \textit{Stabilized convection-diffusion equation}}
\nomenclature{\(f\)}{Source term in the \textit{Stabilized convection-diffusion equation}}

\nomenclature{\(K_I\)}{Loading stress intensity factor for the boundary layer}
\nomenclature{\(\nu\)}{Poisson's ratio}
\nomenclature{\(R_b\)}{Radius of remote boundary layer}
\nomenclature{\(\theta\)}{Angle of remote boundary points to the crack plane}
\nomenclature{\(b\)}{Crack tip opening}
\nomenclature{\(n_d\)}{Multiplicative factor for dislocation velocity}
\nomenclature{\(r\)}{Distance from the crack tip}

\nomenclature{\(D_L^0\)}{Pre-exponential diffusion coefficient}
\nomenclature{\(E_L\)}{Activation energy for lattice diffusion}

\printnomenclature
\end{framed}

\section{Introduction}
\label{Introduction}

Hydrogen has the potential to become a major energy commodity which can enable low- or zero-emission energy use in several of the world's energy sectors, such as power generation, road and rail transport, aviation, sea transport and energy and emission-intensive industries. A major challenge for the adoption of H$_2$ fuel on a large scale is the susceptibility of metallic materials to different hydrogen degradation processes. Additionally, prediction of hydrogen distributions within metals and alloys is crucial in many of these energy applications, such as hydrogen transport or storage as a compressed gas \cite{Olden2012HydrogenSimulations, Askari2006PracticalEmbrittlement} or in hydride-forming materials \cite{Zaika2019ModelPhases}. Similarly, the mitigation of hydrogen-isotopes permeation in nuclear reactors also requires the knowledge of diffusion and solubility properties \cite{Hodille2021ModellingConditions}. Many of these processes involve different phenomena needing a multiphysics framework and coupled electro-chemo-mechanical modelling \cite{Hageman2022AnElectrolytes}.\\

The complex interaction of hydrogen and metals also comprises material degradation phenomena, especially hydrogen embrittlement in steels and other alloys, e.g. nickel, titanium or zirconium alloys. In order to understand hydrogen-related failures, coupled models must be adopted in combination with fracture mechanics analyses; hence, hydrogen transport near a crack tip has been numerically studied by many authors \cite{Krom1999HydrogenTip, DiLeo2013HydrogenDeformations, Diaz2016AMetals, Lufrano1996ModelingTip, Martinez-Paneda2020GeneralisedTips,Turnbull1996ModellingTip, Dadfarnia2014ModelingDislocations}. The first milestone for hydrogen uptake, diffusion and trapping modelling near a remotely loaded crack tip was established by Sofronis and McMeeking in 1989 \cite{Sofronis1989NumericalTip}; from this starting point, various works have been presented that improve this coupled deformation-diffusion crack tip phenomena and incorporate other relevant physical processes. Krom et al. \cite{Krom1999HydrogenTip} extended the analysis from Sofronis and McMeeking \cite{Sofronis1989NumericalTip} by considering the influence of the plastic strain rate and the resulting creation of traps in hydrogen distributions. The role of multiple retention sites with different trap densities and energies can be incorporated by including all the corresponding terms in the mass balance, as first studied by Dadfarnia et al. \cite{Dadfarnia2011HydrogenEmbrittlement}. Works based on this framework usually assume that hydrogen flux is caused by lattice diffusion, but Dadfarnia et al. \cite{Dadfarnia2014ModelingDislocations} also considered a flux contribution to model hydrogen transport by mobile dislocations. Similarly, the common assumption of local equilibrium between lattice and trapped hydrogen can be replaced by a more general kinetic exchange, which was originally assessed for hydrogen transport near a crack tip by Turnbull et al. \cite{Turnbull1996ModellingTip}. These analyses usually adopt conventional von Mises plasticity to describe the constitutive behaviour of the material. However, Mart\'{\i}nez-Pa\~neda et al. \cite{Martinez-Paneda2016StrainTip,Martinez-Paneda2016StrainCracking} adopted more quantitative strain gradient plasticity models, showing that GNDs and plastic strain gradients govern material deformation near the crack tip, which results in a higher hydrogen accumulation. Di Leo and Anand \cite{DiLeo2013HydrogenDeformations} proposed the use of the chemical potential, instead of lattice concentration, as the primary variable for the hydrogen transport governing equation. This approach naturally captures stress effects on hydrogen uptake. The need for realistic boundary conditions in electrochemical hydrogen uptake near a crack tip also inspired the development of generalised entry fluxes by Turnbull and co-workers \cite{Turnbull1996ModellingTip, Martinez-Paneda2020GeneralisedTips} and Hageman and Mart\'{\i}nez-Pa\~neda \cite{Hageman2022AnElectrolytes}.
\\

\begin{figure}[H]
  \makebox[\textwidth][c]{\includegraphics[width=1.0\textwidth]{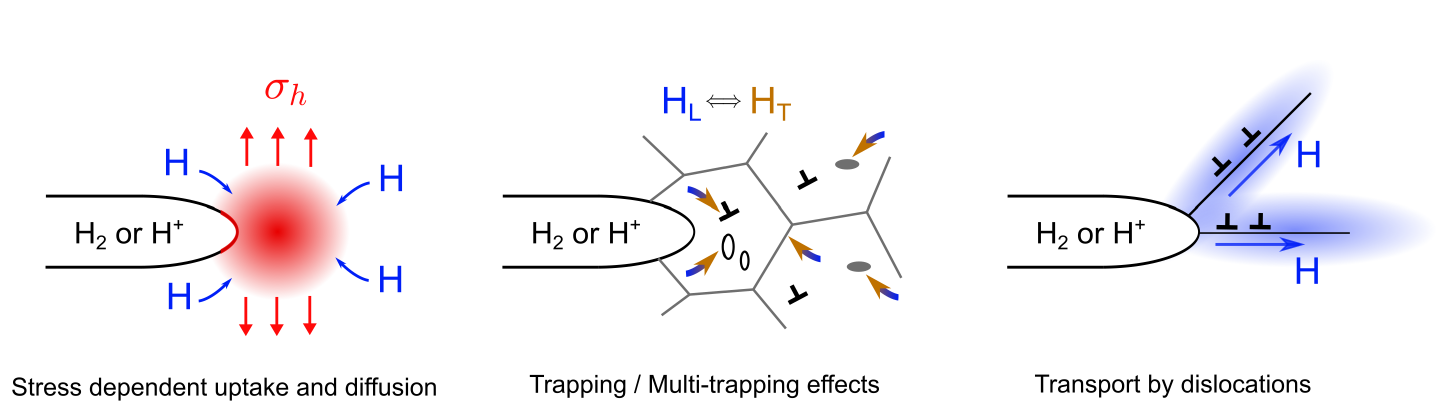}}%
  \caption{Different hydrogen transport phenomena are modelled and implemented in the present work. Hydrostatic stress ($\sigma_h$) influences not only hydrogen diffusion but also uptake on the crack surface. Trapping effects, considering features such as grain boundaries, dislocations, inclusions or voids, are modelled from a local equilibrium or kinetic exchange between lattice hydrogen (L) and trapped hydrogen (T). In addition, hydrogen transport by dislocations assuming a preferred orientation is modelled.}
  \label{fig: transport scheme}
\end{figure}

In the present work, a generalised formulation to describe hydrogen-metal interactions ahead of cracks and other stress concentrators is presented. The theoretical and computational schemes presented incorporate all the key model developments outlined in the previous paragraph and combine them for the first time into a single model. Fig. \ref{fig: transport scheme} outlines some of the key physical processes accounted for. Particular focus is placed on numerical implementation aspects, discussing discretisation, stabilisation and solution strategies. Moreover, all these modelling ingredients are implemented into the commercial finite element package COMSOL Multiphysics, and the models developed are openly shared with the community to maximise progress. The thermodynamic and mechanical theory behind all the implemented transport phenomena is presented in Section \ref{Sec:Theory}, including kinetic and equilibrium trapping, multiple trapping, dislocation transport, stress-dependent hydrogen uptake and generalised entry flux in electrochemical charging. This is followed by COMSOL implementation details in Section \ref{Sec:Comsol}. The results obtained are then presented and discussed in Section \ref{Sec:Results}, and the manuscript ends with concluding remarks in Section \ref{Sec:Concluding remarks}. The hydrogen transport modelling framework here developed can be readily coupled to models that explicitly resolve the cracking process, as elaborated in Part II of this work \cite{Diaz2024AFracture}.

\section{Theory}
\label{Sec:Theory}

\subsection{Modified mass balance}
The seminal work of Sofronis and McMeeking \cite{Sofronis1989NumericalTip} considers two modifications to the classical mass balance equation:
\begin{equation}
\label{Eq. mass balance}
    \frac{\partial C}{\partial t}+\nabla \cdot \mathbf{J}  = 0
\end{equation}
\noindent where $C$ is the total hydrogen concentration and $\mathbf{J}$ a hydrogen flux vector. The first modification is based on the partition of total concentration in lattice and trapped hydrogen:
\begin{equation}
\label{Eq. governing flux not expanded}
    \frac{\partial C_L}{\partial t}+\frac{\partial C_T}{\partial t}+\nabla \cdot \mathbf{J}  = 0
\end{equation}
\noindent where $C_L$ and $C_T$ represent the local concentration of hydrogen in lattice and trapping sites, respectively. Traps are defined as retention sites that delay diffusion \cite{cupertino2023hydrogen}. Different microstructure features can be considered as trapping sites: dislocations, grain boundaries, inclusions, vacancies, etc. In the present two-level approach, all trapping effects are comprised within $C_T$ and the trap energy and density, defined below. Nevertheless, a multi-trapping version of the mass balance is also considered in Section \ref{Sec: multi-trapping}. The second modification includes a hydrogen flux dependent on hydrostatic stress, due to the decrease of chemical potential of lattice sites in tensile regions. The chemical potential $\mu_L$ can be expressed as a function of lattice occupancy, defined as the ratio between the interstitial hydrogen concentration and the density of lattice sites ($\theta_L = C_L/N_L$) \cite{DiLeo2013HydrogenDeformations}:
\begin{equation}
\label{Eq. mu_L vs C_L}
    \mu_L=\mu_L^0+RT \ln{\frac{\theta_L}{1-\theta_L}}-\bar{V}_H\sigma_h
\end{equation}
\noindent where  $\mu_L^0$ is the reference chemical potential, $\bar{V}_H$ the partial molar volume of hydrogen in the host metal, $\sigma_h$ the hydrostatic stress, i.e. the trace of the stress tensor divided by 3, $T$ the temperature and $R$ the constant of gases. In addition, lattice occupancy is assumed to be low ($\theta_L \ll 1$) as most metals have a low hydrogen solubility. Considering trapping and stress features, the classical Fick's second law is modified as follows:
\begin{equation}
\label{Eq. governing eq}
    \frac{\partial C_L}{\partial t}+\frac{\partial C_T}{\partial t}+\nabla \cdot \left(-D_L\nabla C_L+\frac{D_L \Bar{V}_H}{RT} C_L\nabla \sigma_h \right) = 0
\end{equation}

\noindent where $D_L$ is the ideal diffusivity through lattice sites. It must be noted that flux, the term between brackets in Eq. (\ref{Eq. governing eq}), is assumed to occur through lattice sites, i.e. $\mathbf{J}=\mathbf{J}_L$ and cross-site fluxes are neglected; details of a more general diffusion framework can be found in Ref. \cite{Toribio2015ATypes}.\\

Predicting the evolution of $C_L$ by solving the mass balance equation, Eq. (\ref{Eq. governing eq}), is relatively straightforward but requires computing the hydrostatic stress gradient, as discussed in Subsection \ref{Diff-convect-reaction_implementation}. In addition, $C_T$ can be considered a dependent state variable, uniquely defined by $C_L$, or an independent variable, with the latter requiring resolving the term $\partial C_T/\partial t$, i.e. to establish an expression for the time evolution of trapped hydrogen, as discussed below.

\subsection{Evolution of trapped hydrogen}
The relationship between the lattice and trapped hydrogen concentration, $C_L$ and $C_T$, respectively, can be implemented assuming equilibrium or a general kinetic exchange.

\subsubsection{Oriani's equilibrium}
\label{Subsub:theroy_eq}
Thermodynamic equilibrium, termed in this context as Oriani's equilibrium \cite{Oriani1970TheSteel}, is usually assumed between interstitial and trapped hydrogen:
\begin{equation}
\label{eq. equilibrium}
    C_T = \frac{K_T N_T \theta_L}{1+K_T \theta_L}
\end{equation}

\noindent where $K_T$ is the equilibrium constant and $N_T$ is the trap density. Trap occupancy can be defined as the ratio between the hydrogen concentration at traps and the trap density, $\theta_T=C_T/N_T$. The equilibrium constant depends on the binding energy of traps, $E_B$, following an Arrhenius behaviour: $K_T=\exp(E_B/RT)$. Considering the derivation from \cite{Sofronis1989NumericalTip}:
\begin{equation}
\label{eq. equilibrium deriv}
    \frac{\partial C_T}{\partial t}=
    \frac{\partial C_T}{\partial C_L}\frac{\partial C_L}{\partial t} =\frac{C_T(1-\theta_T)}{C_L}\frac{\partial C_L}{\partial t}
\end{equation}

An alternative arrangement of $\partial C_T/\partial C_L$ could be considered \cite{Raina2017AnalysisAlloys} to avoid division by zero for an initial concentration $C_L^0 = 0$: 
\begin{equation}
     \frac{\partial C_T}{\partial t}= \frac{K_T N_T/N_L}{(1+K_TC_L/N_L)^2}\frac{\partial C_L}{\partial t}
\end{equation}

It must be noted that the assumption $\theta_L \ll 1$ is not required to derive the $\partial C_T / \partial t$ term, as shown by Dadfarnia et al. \cite{Dadfarnia2011HydrogenEmbrittlement}. Thus, releasing the $\theta_L \ll 1$ assumption, the reaction term can be expressed as:
\begin{equation}
    \frac{\partial C_T}{\partial t}=\frac{K_T N_T/N_L}{[1+(K_T-1)C_L/N_L]^2}\frac{\partial C_L}{\partial t}
\end{equation}

\noindent Both are typically equivalent because, for the common range of binding energies, $K_T-1 \approx K_T$ \cite{Fernandez-Sousa2020AnalysisFatigue}.
When trap density $N_T$ or temperature are not constant, the chain rule must be applied to consider the influence terms:
\begin{equation}
    \frac{\partial C_T}{\partial t} = \frac{\partial C_T}{\partial C_L}\frac{\partial C_L}{\partial t}+\frac{\partial C_T}{\partial N_T}\frac{\partial N_T}{\partial t}+\frac{\partial C_T}{\partial K_T}\frac{\partial K_T}{\partial T}\frac{\partial T}{\partial t}
\end{equation}
The last term is not here implemented because only isothermal diffusion is simulated, and the term $\partial C_T / \partial N_T$ can be replaced by $ \theta_T$. In addition, the dependence of $N_T$ on plastic strain must be considered in a coupled mechanical-diffusion framework and thus the trapping rate is expressed as:
\begin{equation}
    \frac{\partial C_T}{\partial t}=\frac{K_T N_T/N_L}{[1+(K_T-1)C_L/N_L]^2}\frac{\partial C_L}{\partial t}+\theta_T\frac{d N_T}{d \varepsilon_p}\frac{\partial \varepsilon_p}{\partial t}
\end{equation}
The term derived from Oriani's equilibrium is used to define a non-dimensional operational diffusivity:
\begin{equation}
\label{Eq. Operational D*}
    \bar{D}=\frac{D_L}{D_{eff}}=1+\frac{K_T N_T/N_L}{[1+(K_T-1)C_L/N_L]^2}
\end{equation}
Finally, the mass balance based on the lattice concentration as the primal variable can be expressed as follows assuming equilibrium between trapped and lattice hydrogen:
\begin{equation}
\label{Eq. governing eq with D*}
    \bar{D}\frac{\partial C_L}{\partial t}+\theta_T\frac{d N_T}{d \varepsilon_p}\frac{\partial \varepsilon_p}{\partial t}+\nabla \cdot \left(-D_L\nabla C_L+\frac{D_L \Bar{V}_H}{RT} C_L\nabla \sigma_h \right) = 0
\end{equation}

\subsubsection{McNabb and Foster's equation}
\label{eq:MFtheory}
With a more general validity, the evolution of trapped hydrogen might be expressed through the difference between two terms: a capture term proportional to the amount of lattice hydrogen and the fraction of empty traps and, similarly, a release term depending on the amount of trapped hydrogen and to the fraction of empty interstitial sites. The domain of validity of Oriani's assumption, i.e. local equilibrium, has been discussed for thermal desorption spectroscopy (TDS) \cite{Ebihara2009ASpectra,garcia2024tds} and for hydrogen transport near a crack tip during fast loading \cite{Charles2021EffectTip}. Following the notation by Krom and Bakker \cite{Krom2000HydrogenSteel}, the kinetic exchange is expressed as:
\begin{equation}
\label{eq. MCNabb-Foster}
    \frac{\partial C_T}{\partial t} = \kappa \theta_L N_T(1-\theta_T)-\lambda C_T (1-\theta_L)
\end{equation}
where the trap occupancy $\theta_T$ is defined as $C_T/N_T$ and $\kappa$ and $\lambda$ represent the capture and release constants in frequency units [s$^{-1}$]. Sometimes these constants are expressed as $k$ and $p$ \cite{Turnbull1996ModellingTip}, following the original paper of McNabb and Foster \cite{McNabb1963ASteels}, with different units; in that case, $kC_L$ is equivalent to $\kappa\theta_L$ here. It should be noted that the kinetic expression in Eq. (\ref{eq. MCNabb-Foster}) is derived considering $N_T \ll N_L$ \cite{Krom2000HydrogenSteel}. Additionally, the release term is usually simplified as $\lambda C_T$ due to the low occupancy in lattice sites, i.e. $\theta_L \ll 1$, as common in many alloys.

\subsection{Multi-trapping effects}
\label{Sec: multi-trapping}
The validity of an averaged-trap model, i.e. with a binding energy and trap density that represent all defects, should be better explored. Some authors have reproduced hydrogen accumulation near a crack tip in the presence of different trapping sites \cite{Fernandez-Sousa2020AnalysisFatigue, Isfandbod2021AFracture}. At least, the differentiation between reversible and irreversible traps is useful for embrittlement predictions \cite{Carrasco2019NumericalOverprotection}. Experimentally, TDS spectra enable establishing different trapping sites \cite{Chen2020ObservationPrecipitates,garcia2024tds}; similarly, subsequent permeation transients are used to quantify weak and strong trapping effects \cite{Simoni2021AnSteel}. A multi-trap model was presented by Dadfarnia et al.  \cite{Dadfarnia2011HydrogenEmbrittlement}, based on the $n-$partition of the trapping term for the $n$ number of trapping sites that have been defined:
\begin{equation}
    \frac{\partial C_T}{\partial t}= \sum_{i=1}^{n}\frac{\partial C_T^i}{\partial t}
\end{equation}
Each term in this sum can be derived as in Subsection \ref{Subsub:theroy_eq} if equilibrium is assumed or substituted by Eq. (\ref{eq. MCNabb-Foster}) instead. The generalisation of multiple trapping exchange is also detailed in Ref. \cite{Toribio2015ATypes}.

\subsection{Dislocation transport of hydrogen}
\label{Sec:TheoryDislocationsH}

Some hydrogen embrittlement theories are based on the interaction between hydrogen and dislocations; moreover, some authors have proposed that hydrogen accumulation in the fracture process zone might be enhanced by the transport of hydrogen by dislocations \cite{Tien1976HydrogenDislocations}. Dislocation assistance for hydrogen transport has been experimentally confirmed by Pu and Oi \cite{Pu2019HydrogenSteel} through microprinting observations in austenitic stainless steels. Numerically, due to the mass balance form, this can be regarded as dislocation-assisted convection \cite{Neeraj2012HydrogenNanovoiding}. Dadfarnia et al.  \cite{Dadfarnia2014ModelingDislocations} were the first to model this effect by including a flux term accounting for mobile dislocations that carry out trapped hydrogen:
\begin{equation}
    \mathbf{J^d}=C_T^{m} n_d\mathbf{v^d}
\label{Eq. flux H dislocations}
\end{equation}
\noindent where $C_T^{m}$ is the concentration of hydrogen trapped at mobile dislocations and $\mathbf{v^d}$ is the dislocation velocity vector, which can be modelled mechanistically considering its relationship with plastic strain rate, $\partial \varepsilon_p / \partial t$. In addition, a factor $n_d$ is included multiplying $\textbf{v}^d$ to account for any effect that proportionally enhances dislocation velocity. Here, as in Ref. \cite{Dadfarnia2014ModelingDislocations}, a geometric relation between the trap and dislocation density, $N_T^m$ and $\rho^m$, and the Orowan equation are assumed, yielding the following flux expression:
\begin{equation}
\label{Eq: Jd Orowan}
    \mathbf{J^d}=\theta_T^{m} N_T^{m} \mathbf{v^d} = 
    \theta_T^{m} \frac{\gamma \rho^m}{a} n_d\mathbf{v^d}
    = \theta_T^{m}n_d\frac{\gamma}{b_va} \frac{\partial \varepsilon_p}{\partial t}\mathbf{n}
\end{equation}

\noindent where $\gamma$ is a geometric parameter equal to $\sqrt{2}$ for bcc and $\sqrt{3}$ for fcc crystal structures, $b_v$ the Burgers vector and $a$ the lattice parameter. A unitary vector $\mathbf{n}$ is included to account for dislocation transport direction. Equation (\ref{Eq: Jd Orowan}) assumes that the trap density $N_T^m$ is proportional to the density of mobile dislocations $\rho^m$. 
Thus, $N_T^m$ is here expressed in traps per unit volume, so conversion must be considered for other concentration units.
Assuming that the occupancy of mobile dislocations follows equilibrium, hydrogen flux by dislocations can be expressed as a convective term, i.e. a concentration $C_L$ multiplied by a convection velocity term. 
\begin{equation}
\label{Eq. dislocation flux}
    \mathbf{J^d}=
    C_L \frac{K_T}{K_T C_L + N_L} n_d\frac{\gamma}{b_v a} \frac{\partial \varepsilon_p}{\partial t} \mathbf{n}   
\end{equation}
This convection rearrangement is exploited for implementation, as shown in Subsection \ref{Diff-convect-reaction_implementation}.

\subsection{Stress-dependent boundary conditions}
\label{Sec. Stress-dependent BCs}

Traditionally, Sievert's law has been assumed and thus a Dirichlet constraint is usually imposed over the sample or crack boundary $\mathcal{B}$. For a two-level model in which the primary variable is the hydrogen concentration in lattice sites, a concentration in equilibrium with the environment, $C_{env}$, is fixed:
\begin{equation}
    C_L(\mathcal{B}) = C_{env} = K\sqrt{p_{H_2}}
\end{equation}

\noindent where $K$ is the solubility and $p_{H_2}$ the gaseous hydrogen pressure. However, it must be noted that lattice sites are expanded due to hydrostatic stress, which takes non-zero values on a crack tip surface during loading. The equilibrium condition at the surface must not be expressed in terms of concentration but considering the lattice chemical potential, $\mu_L$, in equilibrium with the gaseous hydrogen chemical potential, $\mu_{H_2}$:
\begin{equation}
    \mu_L(\mathcal{B}) = \frac{1}{2}\mu_{H_2}
\end{equation}
\begin{equation}
\label{Eq:chemical_potential}
    \mu_L^0+RT\ln{\frac{C_L(\mathcal{B})}{N_L}}-\bar{V}_H\sigma_h(\mathcal{B}) = \frac{1}{2} \left(\mu_{H_2}^0 + \ln{\frac{f_{H_2}}{p^0}}\right)
\end{equation}

\noindent where $\mu_L^0$ and $\mu_{H_2}^0$ are the reference potentials, $p^0$ the reference pressure and $f_{H_2}$ the fugacity, a magnitude alternative to pressure that accounts for deviations from ideal gas behaviour \cite{Marchi2007PermeabilityPressures}. Operating in Eq. (\ref{Eq:chemical_potential}), the equilibrium concentration $C_{env}$ is proportional to the square root of fugacity and a stress-dependent term must be added to the boundary condition:  
\begin{equation}
\label{Eq:CL_stress}
    C_L(\mathcal{B})= \frac{N_L}{\sqrt{p^0}}\exp\left(\frac{-\mu_L^0}{RT}\right)\sqrt{f_{H_2}} \exp\left(\frac{\sigma_h \bar{V}_H}{RT}\right)= C_{env} \exp\left(\frac{\sigma_h \bar{V}_H}{RT}\right)    
\end{equation}
It should be noted that stress values, including $\sigma_h$, are obtained at integration points and thus the stress-dependent concentration as a node boundary condition requires extrapolation in finite element codes. This is discussed in subsection \ref{Subsec:bc_implement}.\\

Alternative to concentration-based modelling, the choice of the chemical potential as the primary variable in hydrogen diffusion problems was proposed by Di Leo and Anand \cite{DiLeo2013HydrogenDeformations}. The mass balance, Eq. (\ref{Eq. governing eq with D*}), can be reformulated as:
\begin{equation}
\label{Eq. governing eq with muL}
    \bar{D}\frac{C_L}{RT}\frac{\partial \mu_L}{\partial t}+\bar{D}\frac{C_L}{RT}\bar{V}_H \frac{\partial \sigma_h}{\partial t}+\theta_T\frac{d N_T}{d \varepsilon_p}\frac{\partial \varepsilon_p}{\partial t}+\nabla \cdot \left(-D_L\frac{C_L}{RT}\nabla \mu_L \right) = 0
\end{equation}
where the relationship between $\mu_L$ and $C_L$, i.e. Eq. (\ref{Eq. mu_L vs C_L}), has been considered for $\theta_L<<1$. From this relationship, the lattice concentration is obtained from the dependent variable, $\mu_L$, as:
\begin{equation}
    \label{Eq. C_L vs mu_L}
    C_L = N_L \exp{\left(\frac{\mu_L-\mu_L^0+\bar{V}_H\sigma_h}{RT}\right)}
\end{equation}

The form of Eq. (\ref{Eq. governing eq with muL}) shows two main advantages in comparison to concentration-based governing equations: (i) realistic boundary conditions are easily implemented; and (ii) it is not necessary to compute the gradient of the hydrostatic stress but only the rate of that magnitude. The implementation of this equation based on $\mu_L$ is described in Section \ref{mu_L implementation}.

\subsection{Generalised boundary conditions}
\label{Sec:GeneralisedBCs}

The boundary conditions for modelling hydrogen uptake during electrolytic charging must consider the imbalance between diffusion, absorption and adsorption. Charging and recombination kinetics are governed by the Hydrogen Evolution Reaction (HER), which has been extensively treated in the context of hydrogen permeation through metals \cite{Iyer1989AnalysisProcess, Bockris1971TheHydrogen, Liu2014DeterminationSteel}. The occupancy of surface sites by adsorbed hydrogen atoms defines the coverage magnitude, $\theta_{ad}$, and links absorption and desorption equations; 
\begin{equation}
\label{Eq: Absorption}
    J(\mathcal{B})= k_{abs}\exp\left(\frac{\sigma_h \bar{V}_H}{RT}\right)\theta_{ad}-k_{des}C_L(1-\theta_{ad})
\end{equation}
\begin{equation}
\label{Eq: Adsorption}
    J(\mathcal{B})= k_c (1-\theta_{ad})-k_{r,chem}\theta_{ad}^2-k_{r,elec}\theta_{ad}
\end{equation}

The absorption constant is $k_{abs}$ and the desorption constant is $k_{des}$; the former could be redefined with velocity units substituting $k_{abs}$ by $k_{abs}^*N_L$ \cite{Martinez-Paneda2020GeneralisedTips}. The charging constant, $k_c$, models the Volmer equation so it is influenced by the overpotential and electrolyte pH; the chemical recombination constant, $k_{r,chem}$ represents the Tafel reaction and  $k_{r,elec}$ the electrochemical recombination (Heyrovsky reaction)  \cite{Turnbull2015PerspectivesTrapping}. Turnbull et al. \cite{Turnbull1996ModellingTip} assess the possible simplification of generalised boundary conditions to a constant concentration when diffusion flux is small compared to the absorption ($k_{abs}$) and charging ($k_c$) constants; in that case, the stress-dependent boundary condition can be expressed as in Eq. (\ref{Eq:CL_stress}). The equilibrium concentration $C_{env}$ depends on absorption/desorption constants and the equilibrium coverage: 
\begin{equation}
\label{Eq: C_env electrolyte}
    C_{env}= \frac{k_{abs}^*N_L}{k_{des}}\frac{\theta_{ad}^e}{1-\theta_{ad}^e}
\end{equation}
\noindent where $\theta_{ad}^e$ is the constant surface coverage in equilibrium that can be found as a function of charging and recombination constants \cite{Turnbull1996ModellingTip}. For a more comprehensive description of hydrogen uptake, including the handling of the electrochemical behaviour of the electrolyte, the reader is referred to Refs. \cite{Hageman2022AnElectrolytes,Cupertino-Malheiros2024HydrogenElectrolytes}.

\subsection{Hydrogen-induced softening and dilatation}

Sofronis et al. \cite{Sofronis2001HydrogenAlloys} first proposed a linear form to model hydrogen-induced softening:
\begin{equation}
    \sigma_y = \sigma_{y0}f_s(c)h(\varepsilon_p)
\label{Eq. hardening}
\end{equation}

\noindent where $\sigma_{y0}$ is the initial yield stress in the absence of hydrogen, $h(\varepsilon_p)$ is the strain hardening law, and $f_s(c)$ represents the softening behaviour produced by the local hydrogen concentration expressed as hydrogen atoms per metal atom:
\begin{equation}
    f_s(c)=(\zeta-1)c+1
\label{Eq. H-induced softening}    
\end{equation}
The coefficient $\zeta\le 1$ can capture different softening levels: for $\zeta = 1$, no softening is modelled and the limiting case for $c=1$ H/M results in a yield stress of $\zeta\sigma_{y0}$.

\section{Numerical implementation in COMSOL Multiphysics}
\label{Sec:Comsol}

The generalised formulation presented in Section \ref{Sec:Theory}, which captures all the key hydrogen-material interactions governing crack tip behaviour, is implemented in the finite element package COMSOL Multiphysics. This choice is grounded on the advantages of the COMSOL Multiphysics user environment, which include: (i) an equation-based interface that does not require programming or user subroutines; (ii) the possibility to easily couple different physical processes, without the need for sequential analysis or file writing and reading; and (iii) the option to implement advanced modelling features for hydrogen diffusion through the \texttt{Transport of Diluted Species} module and the customised diffusion-convection-reaction terms. Some of the diffusion processes discussed in Section \ref{Sec:Theory} have been modelled using COMSOL Multiphysics \cite{Bouhattate2011ComputationalTrapping, Li2017AnisotropyDiffusion, Yao2019HydrogenCOMSOL,Sanchez2020DiffusionTemperatures}. Also, based on the work by Hageman and Mart\'{\i}nez-Pa\~neda \cite{Hageman2022AnElectrolytes}, an application note (ID: 116021, \textit{Hydrogen Diffusion in Metals}) has also been presented in the newest version of COMSOL to simulate stress-driven hydrogen uptake and diffusion from an aqueous electrolyte\footnote{See \url{https://www.comsol.com/model/hydrogen-diffusion-in-metals-116021}}. In this Section, the implementation of the generalised formulation described in Section \ref{Sec:Theory} is extensively discussed, presenting a robust numerical framework that can accommodate all relevant physical phenomena governing the behaviour of metals exposed to hydrogen-containing environments.\\

To simultaneously solve the coupled deformation-diffusion problem (i.e., displacement and concentration fields), two `Physics interfaces' are defined in the COMSOL environment: \texttt{Solid Mechanics} and \texttt{Transport of Dilute Species} (\texttt{tds}). Alternatively, the modified Fick's laws for hydrogen transport can also be implemented in Comsol considering a general \texttt{Coefficient Form PDE} interface or through a \texttt{Stabilized Convection-Diffusion Equation}. Case studies evaluated in the present work are 2D plane strain problems, but it must be noted that the \texttt{Coefficient Form PDE} is not adapted to axial symmetry. This limitation can be overcome by using the \texttt{Stabilized Convection-Diffusion Equation}.  It must be noted that two variables from the \texttt{Solid Mechanics} analysis inform the \texttt{tds} equation: hydrostatic stress $\sigma_h$ and equivalent plastic strain $\varepsilon_p$. When damage modelling is not considered and the material constitutive response is assumed to be independent of concentration, hydrogen transport and crack tip mechanics are only weakly coupled. This weakly coupled system can be solved in two ways: (1) through the use of a monolithic, \texttt{Fully Coupled} analysis, which is unconditionally stable, or (2) through a sequential approach (also referred to as staggered or \texttt{Segregated}), where the displacement field is first obtained and then passed as a predefined field to the mass diffusion simulation. The latter is considered more robust but can result in inaccuracies if the time increment is not sufficiently small. The accuracy of the simulation can be improved by adopting what is typically referred to as a `multi-pass' approach (vs `single-pass') where, for a given time increment, multiple iterations over the deformation and diffusion problems are conducted. Both monolithic and segregated approaches are considered here and their limitations and strengths are discussed. 

\subsection{Physics 1: \texttt{Solid Mechanics}}

The solution field in the \texttt{Solid Mechanics} physics interface is the displacement field, $\mathbf{u}$. The balance in linear momentum defines the governing equation. Details of the mechanical problem are not included as these are standard. %In quasi-static conditions, without body forces and for a geometrically nonlinear formulation, the balance can be expressed as:
%\begin{equation}
%    \nabla_0 \cdot (\mathbf{F} \mathbf{S}) = 0
%\end{equation}
%where $\mathbf{F}$ is the deformation gradient tensor determined as $\mathbf{F}=\mathbf{I}+\nabla \mathbf{u}$, being $\mathbf{I}$ the identity matrix. The divergence operator is computed to ther reference configuration ($\nabla_0$) and $\mathbf{S}$ represents the second Piola-Kirchhoff stress. However, hydrostatic stress values to be passed to the diffusion problem are calculated from the Cauchy stress tensor.  The deformation gradient is composed of multiplicative elastic and plastic parts ($\mathbf{F}=\mathbf{F}_e\mathbf{F}_p$) and the Piola-Kirchhoff stress is calculated accordingly. 
Elastic-plastic material behaviour is simulated using von Mises plasticity. Isotropic hardening behaviour is implemented through an analytical hardening expression, $h(\varepsilon_p)$, as follows:
\begin{equation}
    h(\varepsilon_p)=\left( 1+E\frac{\varepsilon_p}{\sigma_{y0}}\right)^N \
\end{equation}
\noindent where $N$ is the strain hardening exponent ($0\leq N \leq 1$) and $E$ is Young's modulus. This power-law hardening expression can also be modelled in COMSOL through a Swift model. To retain generality, we consider a non-linear analysis with large strains and displacements.\\ 

As will be extensively discussed in Section \ref{Sec. Sofronis and Krom benchmark}, special care must be taken when mapping the hydrostatic stress field to accurately compute its gradient in large deformation problems. To ensure an accurate mapping, an additional dependent variable can be created through a \texttt{Weak contribution} in the \texttt{Solid Mechanics} interface:\texttt{(Sh-nojac(-solid.p))*test(Sh)}, where \texttt{test()} is COMSOL's test function for the definition of weak contributions, \texttt{nojac()} is an operator used to prevent the inclusion of terms in the Jacobian, and the hydrostatic stress variable (\texttt{Sh}) is introduced as an auxiliary dependent variable, defined through the pressure (\texttt{solid.p}), since $p=-\sigma_h$.

\subsection{Physics 2: \texttt{Transport of Diluted Species (tds)}}

In this case, the variable to solve for is the lattice concentration, $C_L$. All the terms from the governing equation, the mass balance, can be intuitively implemented since the \texttt{tds} module is designed to model diffusion, convection and reaction terms. Additionally, the conservative convection form includes a convective velocity within the divergence term, which facilitates the definition of the gradient of the hydrostatic stress. As discussed before, and shown below, an accurate description of hydrostatic stress gradients benefits from the definition of $\sigma_h$ as a nodal unknown  \cite{Yan2014AProblems}. Stress fields can also be incorporated using analytical solutions, e.g. assuming a Prandtl field near a crack tip \cite{Turnbull1996ModellingTip}, or with external results using COMSOL's interpolation function from a file containing the stress components and the corresponding coordinates. The latter could be relevant when inputting information from lower scales (e.g., MD or discrete dislocation dynamics calculations). Stabilization methods for convection-diffusion equations are not here discussed in depth but both consistent and inconsistent methods are available as part of the \texttt{tds} module.

\subsubsection{Diffusion-convection-reaction equation}
\label{Diff-convect-reaction_implementation}

The governing equation in the \texttt{tds} interface includes a diffusive term by default, in which a diffusion coefficient, here $D_L$, multiplies the concentration gradient. Convection has to be activated and the conservative form must be chosen so as to include the convective velocity within the divergence;
\begin{equation}
\label{Eq.diffusion-convection-reaction}
    \frac{\partial C_L}{\partial t}+\nabla \cdot (-D_L\nabla C_L+\mathbf{v} C_L) = R_T
\end{equation}

\noindent where $\mathbf{v}$ represents the convection velocity field and $R_T$ the reaction rate that is exploited here to implement trapping effects. The convective velocity is proportional to the hydrostatic stress gradient:  
\begin{equation}
    \mathbf{v} = \frac{D_L \Bar{V}_H}{RT} \nabla \sigma_h
\end{equation}
This velocity can also be used to implement dislocation transport in addition to stress effects. Following Eq. (\ref{Eq. flux H dislocations}):
\begin{equation}
    \mathbf{v} = \frac{D_L \Bar{V}_H}{RT} \nabla \sigma_h + \frac{K_T}{K_T C_L + N_L} N_T^m \mathbf{v^d}
\end{equation}
and this term is here expanded assuming thermodynamic equilibrium between the hydrogen at lattice sites and the hydrogen trapped at mobile dislocations. The mechanistic relationship with the plastic strain rate is captured by accessing the internal variable \texttt{solid.epet} in COMSOL;
\begin{equation}
    \mathbf{v} = \frac{D_L \Bar{V}_H}{RT} \nabla \sigma_h + \frac{K_T}{K_T C_L + N_L} \frac{\gamma}{b_v a} \frac{\partial \varepsilon_p}{\partial t} \mathbf{n}
\end{equation}

All components for the velocity vector must be implemented individually. For example, for the 2D problems, when assuming only stress effects:
\begin{equation}
    v_x = \frac{D_L \Bar{V}_H}{RT} \frac{d\sigma_h}{dx}
\end{equation}
\begin{equation}
    v_y = \frac{D_L \Bar{V}_H}{RT} \frac{d\sigma_h}{dy}
\end{equation}

The auxiliary dependent variable $\sigma_h$ enables the computation of gradients using the in-built differentiation operators \texttt{d(Sh,x)} and \texttt{d(Sh,y)}. On the other side, the reaction rate must include the trapping effect; if Oriani's equilibrium and low lattice occupancy ($\theta_L << 1$) are assumed:
\begin{equation}
    R_T = -\frac{C_T(1-\theta_T)}{C_L}\frac{\partial C_L}{\partial t}
\end{equation}
or alternatively, to avoid division by zero or a very small number at low hydrogen concentrations:
\begin{equation}
\label{Eq. R_T trap}
    R_T = -\frac{K_T N_T/N_L}{(1+K_TC_L/N_L)^2}\frac{\partial C_L}{\partial t}
\end{equation}

Even though some works \cite{Sofronis1989NumericalTip} define an effective operational diffusivity, $D_{eff}$, as previously derived in Section \ref{Sec:Theory}, the \texttt{tds} module in COMSOL Multiphysics does not include the possibility of defining a damping coefficient equal to $\bar{D}=D_L/D_{eff}$ multiplying $\partial C_L / \partial t$, and therefore the reaction-based arrangement described above is the only possibility. However, a \texttt{Stabilized Convection-Diffusion Equation} could be considered as an alternative implementation strategy. In any case, the strain-rate term proposed by Krom et al. \cite{Krom1999HydrogenTip} requires a reaction factor:
\begin{equation}\label{eq:KromTermImpl}
    R_T = -\frac{C_T(1-\theta_T)}{C_L}\frac{\partial C_L}{\partial t}-\theta_T\frac{dN_T}{d\varepsilon_p}\frac{\partial \varepsilon_p}{\partial t}
\end{equation}

Again, the last term can be easily implemented in the reaction term since $\partial \varepsilon_p / \partial t$ is accessed through the variable \texttt{solid.epet} from the Solid Mechanics problem. On the other hand, if equilibrium cannot be assumed and the McNabb and Foster's kinetics equation needs to be implemented:
\begin{equation}
    R_T = -\frac{\partial C_T}{\partial t}
\end{equation}
In this case, the evolution of $C_T$ cannot be directly derived from $C_L$ and an additional PDE is needed. Implementation details of this formulation, requiring an additional COMSOL \texttt{Physics} interface, are given in Section \ref{Sec: Physics PDE McNabb}.\\

A limitation in the use of the \texttt{Transport of Diluted Species} Physics is that the weak expression of the transport equation is built using the spatial frame, i.e. $\textbf{x}$. However, this can be inaccurate when the transport problem is coupled to a finite deformation problem, i.e. when geometric non-linearity is considered. Since version 6.2, COMSOL includes a new specific transport module with a diffusion equation in the material frame $\textbf{X}$: \texttt{Transport in Solids}. The implementation strategy is analogous to the \texttt{Transport of Diluted Species} and follows Eq. \ref{Eq.diffusion-convection-reaction}, with two minor differences: (i) the reaction term is named as a source term and (ii) an external flux is needed instead of a convective term:
\begin{equation}
    \frac{\partial C_L}{\partial t}+\nabla_\textbf{X} \cdot (-D_L\nabla_\textbf{X} C_L+\Gamma_{ext}) = R_T
\end{equation}
where the external flux now includes $C_L$ and the hydrostatic stress gradient is determined using the material gradient, i.e. with respect to $\textbf{X}$:

\begin{equation}
    \Gamma_{ext,X} = C_L\frac{D_L \Bar{V}_H}{RT} \frac{d\sigma_h}{dX}
\end{equation}
\begin{equation}
    \Gamma_{ext,Y} = C_L\frac{D_L \Bar{V}_H}{RT} \frac{d\sigma_h}{dY}
\end{equation}

\subsubsection{Boundary conditions for Transport of Diluted Species}
\label{Subsec:bc_implement}

The stress-dependency of hydrogen uptake emerges from the thermodynamic equilibrium between lattice and environment chemical potentials, as shown in Section \ref{Sec. Stress-dependent BCs}. The concentration expression from Eq. (\ref{Eq:CL_stress}) is implemented as a Dirichlet boundary condition \cite{Diaz2016AMetals,Martinez-Paneda2016StrainTip}. However, the displacement problem must be unaffected by this boundary condition and thus the default option constraining \texttt{All physics} is substituted by a reaction term only applied to the individual $C_L$ dependent variable. Additionally, boundary conditions based on the input flux must also consider the stress effects, as stress-driven flux must also be taken into account in the balance between diffusion, absorption and desorption kinetics \cite{Turnbull1996ModellingTip}. Thus, the convective term that has been implemented in COMSOL must also be activated for the flux boundary condition:
\begin{equation}
    \mathbf{n}\cdot (\mathbf{J}+\mathbf{v}C_L)=J_{in}
\end{equation}

\noindent where the value $J_{in}$ is entered as a general inward flux. Following the HER previously introduced, $J_{in}$ depends on an additional independent variable $\theta_{ad}$, i.e. the surface coverage of adsorbed hydrogen. The system defined by Eqs. (\ref{Eq: Absorption}) and (\ref{Eq: Adsorption}) could be implemented as an additional \texttt{PDE physics} interface, but $\theta_{ad}$ can be directly solved by equating both expressions and solving the resulting second-order equation. The consideration of convective terms in the input surface flux is not only required for the generalised boundary conditions described above, but also when insulated surfaces are modelled, i.e. when $J_{in}=0$.

\subsubsection{Chemical potential-based equation }
\label{mu_L implementation}

In implementing the transport equation based on the chemical potential as the primal variable, Eq. (\ref{Eq. governing eq with muL}), one should note that the \texttt{tds} module does not include a capacity or damping term that multiplies the rate of the primary variable. Therefore, the \texttt{Stabilized convection-diffusion equation} interface is considered:
\begin{equation}
    d_a\frac{\partial \mu_L}{\partial t}+\nabla \cdot (-c_d\nabla \mu_L) = f
\end{equation}
where the damping ($d_a$), diffusion ($c_d$) and source ($f$) terms are defined following Eq. (\ref{Eq. governing eq with muL}) as:
\begin{equation}
    d_a = \bar{D}\frac{C_L}{RT} \, ; \,\,\,\,\,\, c_d = D_L\frac{C_L}{RT} \, ; \,\,\,\,\,\, f = -\bar{D}\frac{C_L}{RT}\bar{V}_H \frac{\partial \sigma_h}{\partial t}-\theta_T\frac{d N_T}{d \varepsilon_p}\frac{\partial \varepsilon_p}{\partial t}
\end{equation}
In this case, the hydrostatic stress can be stored directly as a variable, i.e. \texttt{Sh=-solid.p}, and the spatial time derivative is computed considering the deformed mesh:  $\partial \sigma_h/\partial t = $\texttt{d(Sh,TIME) - d(Sh,x)*d(x,TIME) - d(Sh,y)*d(y,TIME)}. More details of this implementation are given in Ref. \cite{Diaz2024ExplicitMetals}.

\subsection{Physics 3: Stabilized Convection-Diffusion Equation}
\label{Sec: Physics PDE McNabb}

Only when trapping occupancy is not directly obtained from $C_L$, i.e. for the kinetic McNabb and Foster formulation, an additional PDE must be considered. Analytical approximations were proposed by Benannoune et al. \cite{Benannoune2018NumericalEquation} to circumvent the need for this additional degree of freedom, but these are not valid in all regimes and require the use of very small increments to ensure accuracy \cite{Charles2021NumericalValidity}. Moreover, the implementation of the kinetic problem in COMSOL is straightforward using the \texttt{Stabilized Convection-Diffusion Equation} module to solve $C_T$ and then access $\partial C_T / \partial t$ as a reaction term from the \texttt{tds} module. Two options are possible to model the reaction term when kinetic trapping is considered. 

\subsubsection{Option 1: Kanayama et al. (2009)}

Following Kanayama et al. \cite{Kanayama2009ReconsiderationFormulation}, the trapping rate can be extended, following the chain rule, into two terms: a kinetic term derived from McNabb and Foster's original formulation, and a term depending on the creation of traps. The latter is equivalent to the strain-rate term by Krom and co-workers \cite{Krom1999HydrogenTip}. The reaction term would then read:
\begin{equation}
    \frac{\partial C_T}{\partial t} = N_T\frac{\partial \theta_T}{\partial t} + \theta_T\frac{\partial N_T}{\partial t}
\end{equation}
and the additional PDE to resolve the kinetics of trapping is based on the additional dependent variable $\theta_T$; 
\begin{equation}
\label{Eq.: McNabb Foster occupancy}
    \frac{\partial \theta_T}{\partial t} = \kappa \theta_L(1-\theta_T)-\lambda \theta_T
\end{equation}
However, Charles et al.  \cite{Charles2021EffectTip} demonstrated that following this approach, trap creation is accounted for twice. This problem arises because for the derivation of Eq. (\ref{Eq.: McNabb Foster occupancy}), a constant $N_T$ was considered.

\subsubsection{Option 2: Charles et al. (2021)}
\label{SEc:Option2_Charles}

An alternative option that circumvents the inconsistency highlighted in the previous version is to adopt a reaction PDE with $C_T$ as the unknown variable. In this case, Krom et al. \cite{Krom1999HydrogenTip} strain rate term does not need to be explicitly modelled as the influence of the strain rate naturally emerges from the increase in $N_T$ within the kinetic trapping term. The McNabb and Foster equation is then expressed following as \cite{Charles2021EffectTip}:
\begin{equation}
    \frac{\partial C_T}{\partial t} =  \kappa \theta_L(N_T-C_T)-\lambda C_T
\end{equation}
and it is rearranged to be implemented through the \texttt{Stabilized Convection-Diffusion Equation} available in COMSOL with the corresponding damping, absorption and source terms:
\begin{equation}
    \frac{\partial C_T}{\partial t}+(\kappa \theta_L+\lambda) C_T=  \kappa \theta_L N_T
\end{equation}

\noindent where an absorption term is grouped as $\kappa \theta_L+\lambda$, a source term as $\kappa \theta_L N_T$ and the damping coefficient is equal to one. All other coefficients are equal to zero. \\

When this kinetic formulation is implemented, the initial state of traps also influences the system evolution. If pre-charging is simulated, i.e. $C_L = C_L^0$, the trap hydrogen concentration is here assumed to be initially in equilibrium:
\begin{equation}
    C_T^0 = N_T^0\frac{K_T C_L^0}{K C_L^0 + N_L}
\end{equation}

\section{Validation and discussion}
\label{Sec:Results}

\begin{figure}[H]
\makebox[\linewidth][c]{%
        \begin{subfigure}[b]{0.4\textwidth}
                \centering
                \includegraphics[scale=0.7]{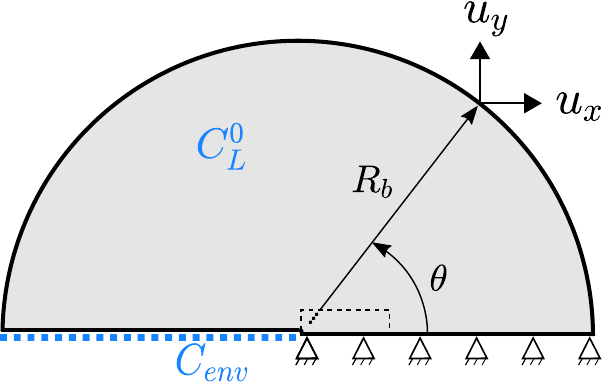}
                \caption{}
                \label{Fig:scheme}
        \end{subfigure}
        \begin{subfigure}[b]{0.4\textwidth}
                \raggedleft
                \includegraphics[scale=0.7]{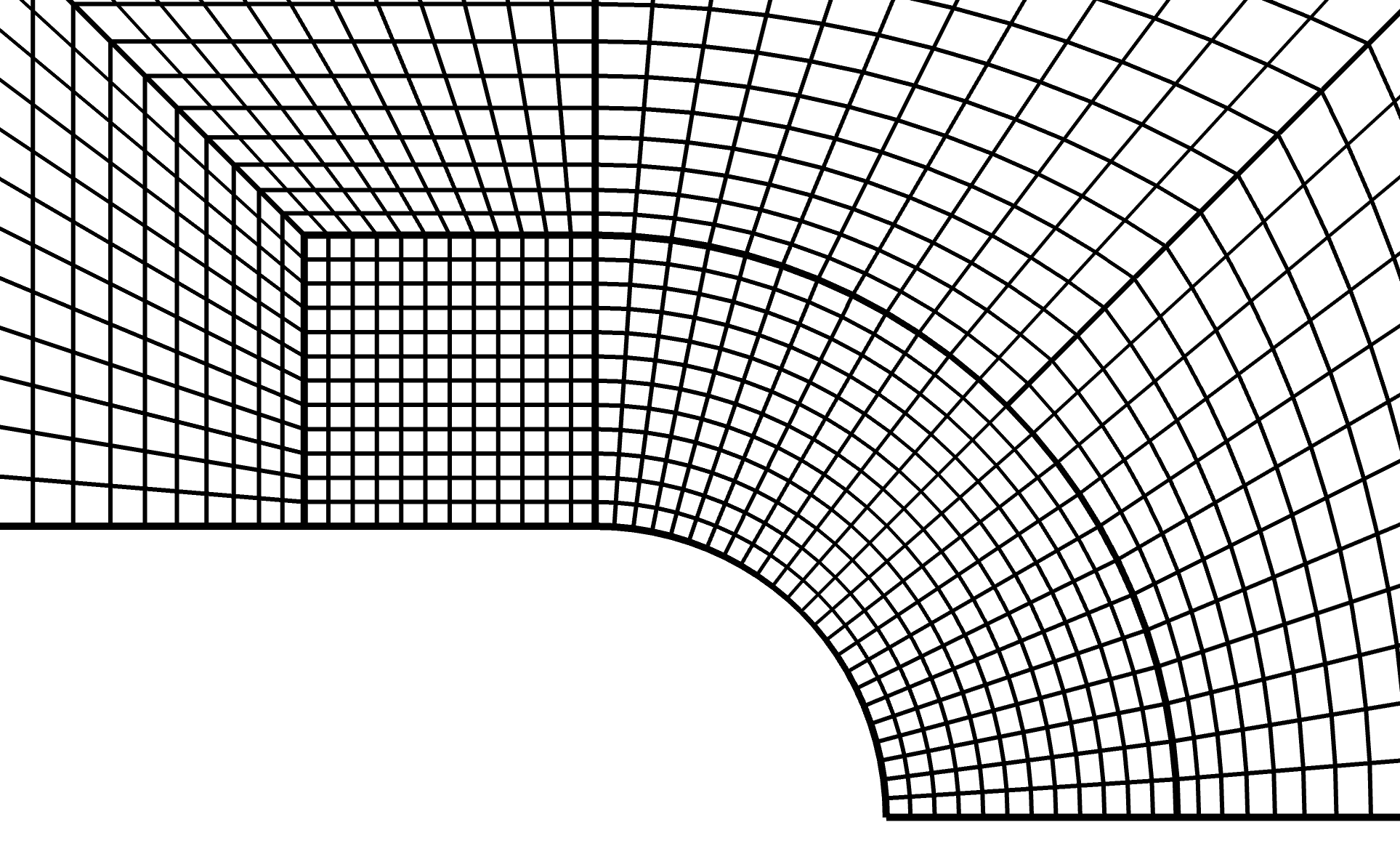}
                \caption{}
                \label{Fig:mesh}
        \end{subfigure}}
        \caption{Boundary layer model used in all simulations: (a) schematic of the geometry and boundary conditions; and (b) detail of the finite element mesh near the crack tip.}
        \label{fig: Multitrap_CL}
\end{figure}

In this Section, the ability of the modelling framework to capture the hydrogen transport phenomena previously described is validated by considering different benchmark problems from the literature. In addition, the influence of discretization schemes, geometrical nonlinearities and the solver choice are discussed. For all the results computed in this Section, a boundary layer approach is considered and plane strain conditions are assumed. The displacements, $u_x$ and $u_y$, applied on the remote boundary, $R_b$, are proportional to the stress intensity factor in mode I, $K_I$. 
\begin{equation}
    u_x(R_b,\theta)=K_I \frac{1+\nu}{E}\sqrt{\frac{R_b}{2\pi}}\cos\left(\frac{\theta}{2}\right)\left[2-4\nu+2 \sin^2{\left(\frac{\theta}{2}\right)}\right]
\end{equation}
\begin{equation}
    u_y(R_b,\theta)=K_I \frac{1+\nu}{E}\sqrt{\frac{R_b}{2\pi}}\sin\left(\frac{\theta}{2}\right)\left[4-4\nu-2 \cos^2{\left(\frac{\theta}{2}\right)}\right]
\end{equation}
where $\theta$ is the angle of each point in the remote boundary with respect to the crack plane and $\nu$ is Poisson's ratio. Symmetry conditions are considered. Figure \ref{Fig:scheme} shows the scheme of the boundary layer and the applied remote displacements. This is the boundary value problem employed in all the literature used for validation. Hence, the focus is on quantifying crack tip behaviour, but the model is of course also applicable to other configurations. For the sake of generality, \ref{App:TDS} describes the application of the present numerical framework to the modelling of thermal desorption spectroscopy (TDS) experiments. \\ 

%\usepackage{graphicx}
%\usepackage{array}
%\usepackage{setspace}

\begin{table}[]
\centering
\caption{Overview of the different transport phenomena considered in the numerical framework presented, including related literature for benchmarking and validation, and details of the implementation strategy, spanning (a) drifted diffusion, (b) trapping, (c) uptake and (d) modified hardening.}
\label{Table: benchmark}
\renewcommand{\arraystretch}{1.4} % Adjust row height for better readability
\resizebox{\textwidth}{!}{%
\begin{tabular}{>{\raggedright\arraybackslash}p{4cm} >{\raggedright\arraybackslash}p{4cm} >{\raggedright\arraybackslash}p{8cm}}
\hline
\textbf{Phenomena} & \textbf{Benchmark} for validation and other references & \textbf{Implementation strategy} \\ \hline
Stress-assisted diffusion and hydrogen trapping & \textbf{Sofronis \& McMeeking} \cite{Sofronis1989NumericalTip} & (a) Convection velocity proportional to $\nabla \sigma_h$ \newline
(b) Reaction term equal to $-\partial C_T/\partial t$ (assuming equilibrium) \newline
(c) Constant concentration as a boundary condition for hydrogen uptake \\ \hline
Depletion of lattice sites for fast creation of traps during plastic deformation & \textbf{Krom et al.} \cite{Krom1999HydrogenTip} & (b) Reaction term including also the contribution of $\partial \varepsilon_p/\partial t$ \\ \hline
Hydrogen trapping in multiple defects & \textbf{Dadfarnia et al.} \cite{Dadfarnia2011HydrogenEmbrittlement} & (b) Reaction term comprising all the trapping contributions ($\sum_i\partial C_T^i/\partial t$) \\ \hline
Hydrogen transport by dislocations & \textbf{Dadfarnia et al.} \cite{Dadfarnia2014ModelingDislocations}& (a) Convection velocity including a term proportional to the dislocation velocity \\ \hline
Kinetic trapping without equilibrium assumptions & Turnbull et al. \cite{Turnbull1996ModellingTip},  Mart\'{\i}nez-Pañeda et al. \cite{Martinez-Paneda2020GeneralisedTips}, \textbf{Charles et al.} \cite{Charles2021EffectTip} & (b) Reaction term determined by solving an additional PDE for $C_T$ based on trapping-detrapping kinetics \\ \hline
Stress influence on hydrogen uptake & \textbf{Di Leo \& Anand} \cite{DiLeo2013HydrogenDeformations}, Díaz et al. \cite{Diaz2016AMetals}, Mart\'{\i}nez-Pañeda et al. \cite{Martinez-Paneda2016StrainTip}, Díaz et al.  \cite{Diaz2024ExplicitMetals}& If the governing PDE is based on $C_L$: \newline
(c) an exponential term considering $\sigma_h$ must be included in the BC \newline
If the governing PDE is based on $\mu_L$: \newline
(c) a constant chemical potential is fixed as the BC \\ \hline
Hydrogen uptake from electrochemical processes & Turnbull et al. \cite{Turnbull1996ModellingTip}, \textbf{Mart\'{\i}nez-Pañeda et al.} \cite{Martinez-Paneda2020GeneralisedTips} & (c) A generalised flux that considers the adsorption/absorption imbalance is established as the BC. Stress effects are also included in the absorption constant. \\ \hline
Hydrogen-modified hardening behaviour & Lufrano et al. \cite{Lufrano1998ElastoplasticallyEmbrittlement}, \textbf{Kotake et al.} \cite{Kotake2008TransientLoading}& (d) Including a phenomenological law for softening as a function of local hydrogen concentration  \\ \cline{1-3}
\end{tabular}%
}
\end{table}

For most case studies, a concentration $C_{env}$ is fixed as a boundary condition in the crack surface and also as the initial condition $C_L^0$. However, insulated or flux boundary conditions are also assessed. The mesh consists of 6,646 elements and is particularly refined near the crack tip, where the characteristic element length is approximately \SI{0.4}{\micro\meter}. The influence of the finite element discretisation is assessed in Section \ref{Sec: Validation multitrapping}.  The relative tolerance is fixed to $10^{-4}$ and a Backward Differentiation Formula (BDF) is chosen as the implicit solver because the method shows a robust stability. In addition, a \texttt{Free} time stepping is selected so the solver automatically controls the increment size depending on the error estimates and the tolerance. Unless otherwise stated, the discretization considers cubic and quadratic Lagrange shape functions for displacement and concentration degrees of freedom, respectively.\\

A summary of all the physical phenomena considered, the relevant references for benchmarking and comparison, and the implementation strategy adopted are given in Table \ref{Table: benchmark}. We begin by validating the implementation of the two-level model considering the influence of hydrostatic stresses and trapping, based on Oriani's equilibrium (Case 1, Section \ref{Sec. Sofronis and Krom benchmark}). This case study evaluates the first two phenomena listed in Table \ref{Table: benchmark}: (i) stress-assisted diffusion and hydrogen trapping, and (ii) depletion of lattice sites for fast creation of traps during plastic deformation. Subsequently, in Section \ref{Sec: Validation multitrapping} (Case 2), the scenario where multiple traps are considered is assessed. Case 3 (Section \ref{Sec:Hdislocations}), considers the transport of hydrogen through dislocation motion. This is followed by Case 4, in Section \ref{Sec:MFresults}, where the kinetic trapping model of McNabb and Foster is assessed. Section \ref{Sec:ChemPotResults} examines the chemical potential-based implementation and the definition of appropriate boundary conditions at the surface (Case 5). This analysis spans two of the phenomena listed in Table \ref{Table: benchmark}: (i) stress influence on hydrogen uptake, and (ii) hydrogen uptake from electrochemical processes. Finally, the last case study (Case 6), addresses the implementation of hydrogen-induced softening (Section \ref{Sec:Hsoftening}).\footnote{All the COMSOL models employed are made freely available at \url{https://mechmat.web.ox.ac.uk/}.}

\subsection{Case 1: Stress and trapping influence considering Oriani's equilibrium}
\label{Sec. Sofronis and Krom benchmark}

The two-level approach, extensively adopted to reproduce hydrogen accumulation near a crack tip, is here implemented and validated by comparing present results with those from Sofronis and McMeeking  \cite{Sofronis1989NumericalTip}, i.e. without considering the strain rate factor, and with results from Krom et al.  \cite{Krom1999HydrogenTip}, i.e. accounting for the influence of trap creation rates. As in the original works, the boundary layer outer radius is chosen to be $R_b=0.15$ m and the initial crack tip opening equals $b_0 =$\SI{10}{\micro\meter}. A ramp load is considered up to a final value of $K_I$ = 89.7 MPa$\sqrt{\text{m}}$ at 130 s; i.e. a 0.69  MPa$\sqrt{\text{m}}$/s loading rate. Krom et al. \cite{Krom1999HydrogenTip} simulated a slightly lower value, $K_I$ = 89.2 MPa$\sqrt{\text{m}}$, but the differences are negligible. The material parameters, aimed at reproducing the behaviour of iron-based materials, are given in Table \ref{Tab:mat_Sofronis}. The evolution of the density of trapping sites is modelled, as in Ref. \cite{Sofronis1989NumericalTip}, following the experimental fitting by Kumnick and Johnson \cite{Kumnick1980DeepIron}:
\begin{equation}
\label{Eq.Kumnick}
    \log N_T = 23.26-2.33\exp{(-5.5\varepsilon_p)}
\end{equation}

\begin{figure}[H]
        \begin{subfigure}[h]{1\textwidth}
                \centering                \includegraphics[scale=0.9]{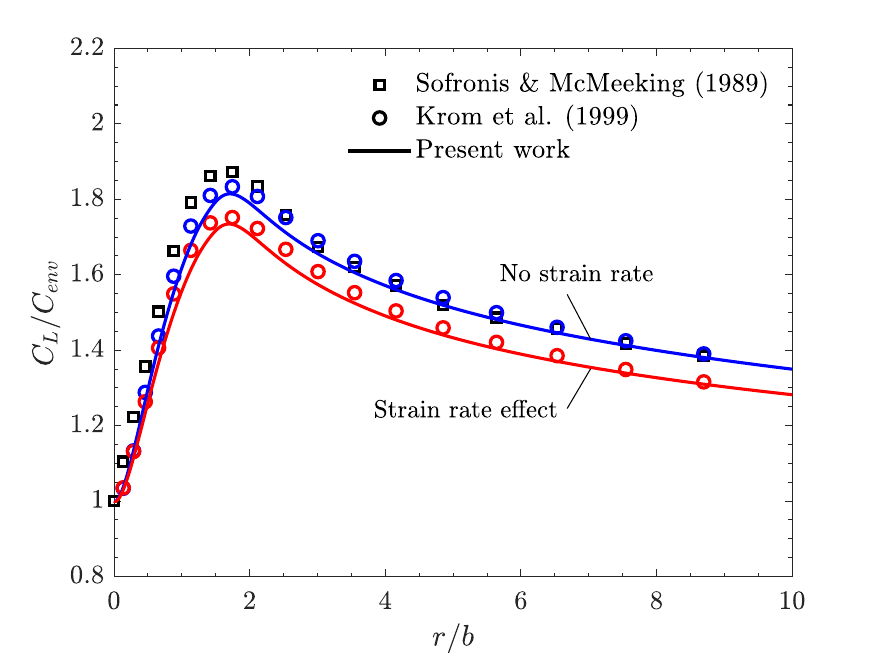}
                \caption{}
                \label{fig:Fig_SM_Krom}
        \end{subfigure}
        \vspace{0.5cm}\\
        \begin{subfigure}[h]{1.1\textwidth}
                \centering               \includegraphics[scale=0.75]{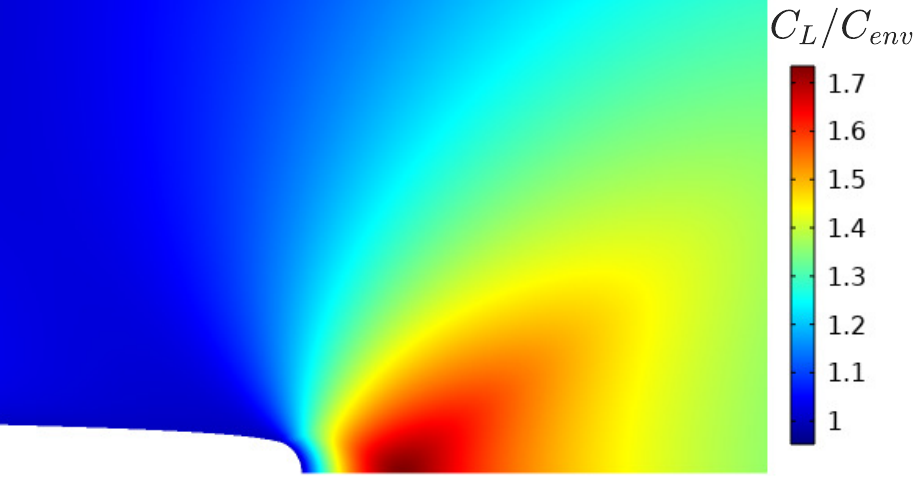}
                \caption{}
                \label{fig:Fig_Krom_Comsol}
        \end{subfigure}       
        \caption{Capturing the influence of stresses and trapping. (a) Comparison of predictions of lattice hydrogen concentration ahead of the crack tip by the present implementation and the works of Sofronis and McMeeking \cite{Sofronis1989NumericalTip} and Krom et al. \cite{Krom1999HydrogenTip}; (b) Contours for normalised hydrogen concentration at lattice sites, $C_L/C_{env}$. The contours in (b) correspond to the maximum loading rate (i.e., after 130 s) and consider the strain rate effect.}
\label{fig: Sofronis and Krom validation}
\end{figure}

\begin{figure}[H]
  \makebox[\textwidth][c]{\includegraphics[width=0.8\textwidth]{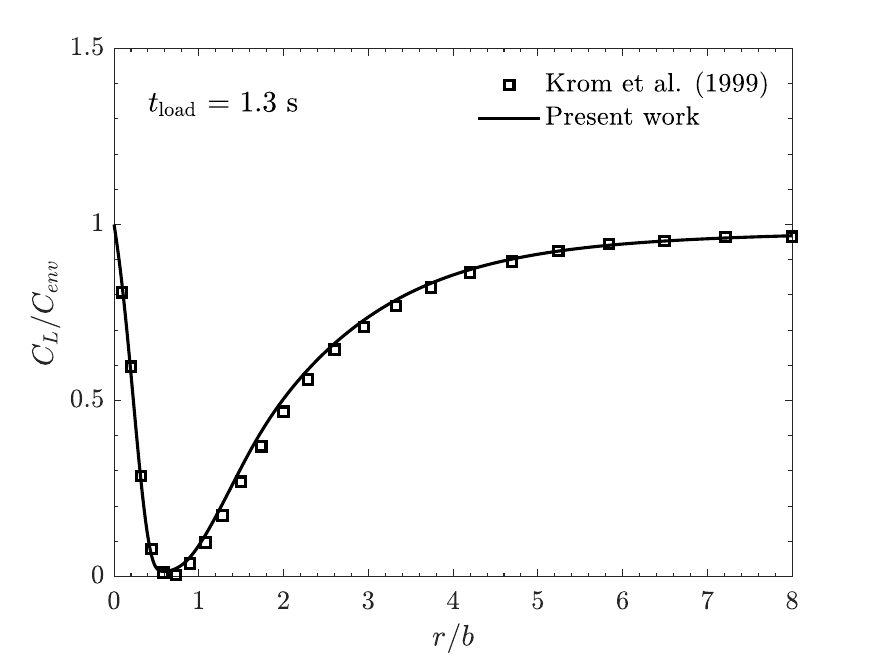}}%
  \caption{Validation of the strain rate effect at high loading rates. Comparison between lattice hydrogen concentrations predicted ahead of the crack tip by the present implementation and the work by Krom et al. \cite{Krom1999HydrogenTip}.}
  \label{fig:Fig_Krom_1s}
\end{figure}

\begin{table}[H]
\centering
\caption{Parameters for the validation of the two-level modelling the role of hydrostratic stresses and trapping, based on Oriani's equilibrium (Case 1), following the works by Sofronis and McMeeking \cite{Sofronis1989NumericalTip} and Krom et al. \cite{Krom1999HydrogenTip}.}
\label{Tab:mat_Sofronis}
   {\tabulinesep=1.2mm
   \makebox[\textwidth][c]{\begin{tabu} {cccccccccccccc}
       \hline
$E$  & $\nu$ & $\sigma_{y0}$ & $N$ & $T$ &   \\ \hline
207 & 0.3  & 250 & 0.2 & 300  \\
(GPa) & (-) &  (MPa) & (-) & (K)    \\ \hline
$D_L$  & $\bar{V}_H$ & $N_L$ & $C_{env}, C_L^0$ & $E_B$\\ \hline
1.27$\times$10$^{-8}$ & 2$\times$10$^{-6}$  & 5.1$\times$10$^{29}$ & 2.084$\times$10$^{21}$ & 60 \\
(m$^2$/s)  & (m$^3$/mol) & (sites/m$^3$) & (atoms/m$^3$) & kJ/mol\\ \hline
        \end{tabu}}}
\end{table}
It must be noted that hydrogen concentrations variables are considered in mol/m$^3$ units within the governing equations. Therefore, $C_{env}$, $C_L^0$, $N_L$ and $N_T$ are divided by the Avogadro constant to convert m$^{-3}$ into mol/m$^3$ units. Relevant results are provided in Fig. \ref{fig: Sofronis and Krom validation}, illustrating finite element predictions of crack tip lattice hydrogen content. The normalised lattice hydrogen concentration ahead of the crack tip is given in Fig. \ref{fig:Fig_SM_Krom}, with the x-axis representing the distance to the crack tip, $r$, normalised by the crack tip opening, $b$, while contours of normalised concentration are given in Fig. \ref{fig:Fig_Krom_Comsol}. The results provided in Fig. \ref{fig:Fig_SM_Krom}, attained for a loading rate of 130 s (when the maximum $K_I$ is reached), demonstrate that the consideration of the plastic strain rate effect shifts the peak of hydrogen in lattice sites towards lower concentrations, because the dynamic creation of traps redistributes hydrogen. Results with and without this strain rate term agree with those from the original references \cite{Sofronis1989NumericalTip, Krom1999HydrogenTip}, validating the present implementation. Further verification is achieved by comparing the outcome of a fast experiment, where the load is applied in only 1.3 seconds, as in Ref. \cite{Krom1999HydrogenTip}. As shown in Fig. \ref{fig:Fig_Krom_1s}, this results in a total lattice depletion, as predicted by both the present model and the study by Krom et al. \cite{Krom1999HydrogenTip}. Once again, a very good quantitative agreement is attained. 

The applied load considered, $K_I$ = 89.2 MPa$\sqrt{\text{m}}$, results in a highly deformed crack tip and significant blunting, as can be seen in Fig. \ref{fig:Fig_Krom_Comsol}. The crack tip opening displacement at the maximum load is found to be 4.5 times the initial tip diameter, i.e. $b=4.5b_0$, whereas Sofronis and McMeeking \cite{Sofronis1989NumericalTip} report $b=5b_0$ and Krom et al. \cite{Krom1999HydrogenTip} $b=4.7b_0$. This could explain the small deviations in lattice hydrogen distribution observed in Fig. \ref{fig:Fig_SM_Krom} (note the x-axis normalisation). More importantly, for the high levels of plastic deformation attained in this boundary value problem, discretisation-dependent errors are found as a result of spurious stress distributions. Quasi-incompressible behaviour during high plastic straining induces volumetric locking \cite{Chen2020CrackModel} and thus the influence of discretization order is assessed. \\

When $\sigma_h$ is stored via a weak contribution, a discretization order must be chosen for three field variables ($\textbf{u}$, $\sigma_h$, $C_L$). The following scenarios are considered: (i) All fields discretized with linear elements: p1 ($\textbf{u}$, $\sigma_h$, $C_L$); (ii) Quadratic-order elements for the displacements and the hydrostatic stress, and linear elements for the lattice hydrogen concentration: p2 ($\textbf{u}$, $\sigma_h$), p1 ($C_L$);(iii) Quadratic-order elements for the displacements and linear elements for the hydrostatic stress and the lattice hydrogen concentration: p2 ($\textbf{u}$), p1 ($\sigma_h$, $C_L$)

\begin{figure}[H]
        \begin{subfigure}[h]{1\textwidth}
                \centering                \includegraphics[scale=0.8]{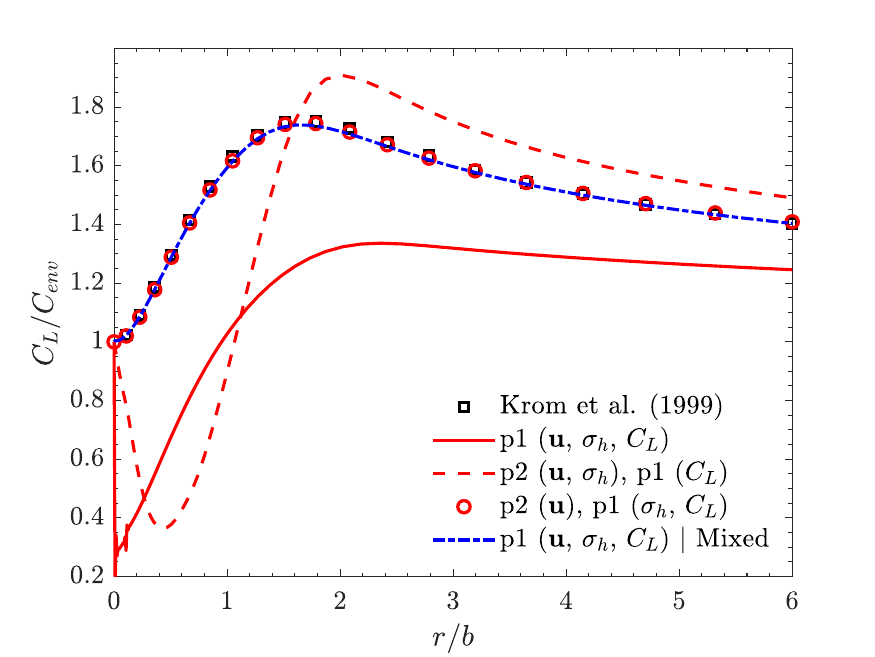}
                \caption{}
                \label{fig: Weak_contribution}
        \end{subfigure}\\
        \begin{subfigure}[h]{1\textwidth}
                \centering               \includegraphics[scale=0.8]{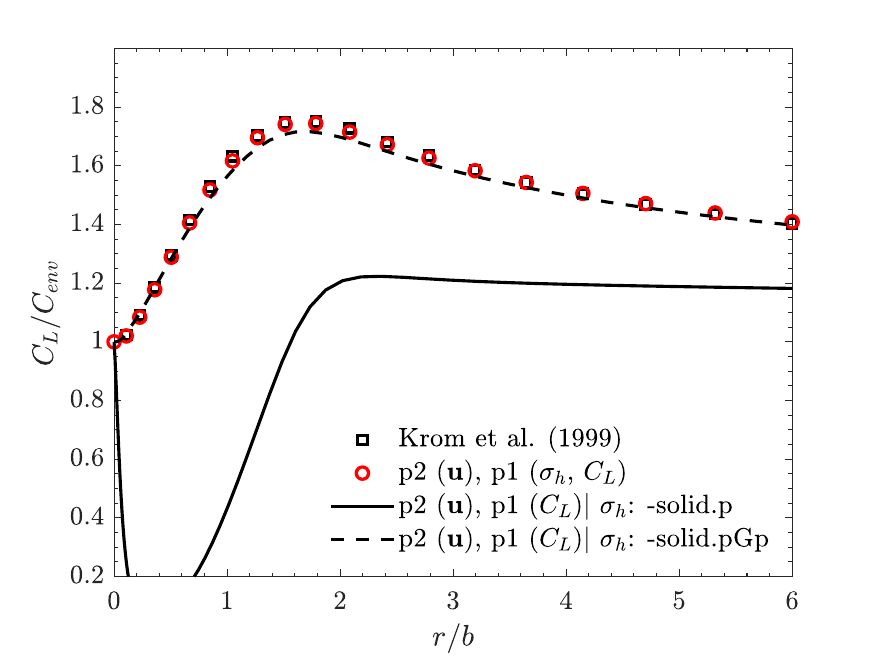}
                \caption{}
                \label{fig: Sh_variable}
        \end{subfigure}       
        \caption{Influence of the discretization scheme on the lattice hydrogen distribution ahead of the crack tip. The results aim at assessing: (a) The role of different p-discretization approaches for the three unknown variables: the displacement field $\mathbf{u}$, the hydrostatic stress $\sigma_h$, and the lattice hydrogen concentration $C_L$; (b) The role of $\sigma_h$ storage approaches in the predictions of lattice hydrogen concentration ahead of the crack tip.}
\label{fig: Multitrap_Ctotal_linear}
\end{figure}

The results obtained are shown in Fig. \ref{fig: Weak_contribution}. It can be seen that the choice of first (p1) order elements for the displacement field produces a spurious non-physical decrease in hydrogen concentration at lattice sites near the crack tip. This is caused by numerical noise in stresses and therefore in error accumulation in the $\nabla \sigma_h$ calculation. An artificial decrease in hydrogen concentration is also observed for a quadratic discretization of displacements and hydrostatic stress. Only a higher order of displacements in comparison to the hydrostatic stress, e.g. p2 ($\textbf{u}$), p1 ($\sigma_h$, $C_L$),  yields correct distributions, regardless of the discretization of $C_L$. These problems were not observed for lower-order elements in ABAQUS \cite{Diaz2016CoupledAnalogy} because it uses a B-bar method or selective reduced integration \cite{deSouzaNeto2005F-bar-basedBenchmarking}, i.e. full integration for deviatoric strains but reduced integration for volumetric strains.\\ 

An alternative solution to avoid these spurious stress oscillations while using low-order elements is to adopt mixed formulations typically used for nearly incompressible materials. If a pressure formulation is activated within the \texttt{Linear Elastic Material} node in COMSOL Multiphysics, an auxiliary pressure $p_w$ is added as an additional variable to the problem and the noise in stress distributions disappears even for the p1 ($\textbf{u}$, $\sigma_h$, $C_L$) discretization, as shown in Fig. \ref{fig: Weak_contribution}. The use of a mixed formulation in stress-assisted hydrogen diffusion was also recently shown to prevent volumetric locking in Ref. \cite{LopesPinto2024SimulationElements}. Thus, it can be concluded that the highly deformed elements at the crack tip are suffering from volumetric locking and this can be prevented: (i) by increasing the discretization order for the displacement field; or (ii) by considering a mixed formulation with the hydrostatic stress (or the pressure) as an additional degree-of-freedom. \\

It is also important to note that when the hydrostatic stress is stored using a conventional variable instead of using a weak contribution, $\sigma_h$ is not accurately mapped from the pressure value, i.e. from \texttt{-solid.p}. In this case, the discretization scheme p2 ($\textbf{u}$), p1 ($C_L$) results in a spurious stress and the corresponding incorrect hydrogen distribution (Fig. \ref{fig: Sh_variable}). However, the built-in evaluation operator at Gauss points transforms the pressure value into a smooth field \texttt{-solid.pGp}, which solves the spurious gradient and accurately predicts $C_L$ distributions. The latter strategy circumvents the need of an auxiliary dependent variable.

\subsection{Case 2: Multi-trapping effects}
\label{Sec: Validation multitrapping}

The influence of multiple trap types (grain boundaries, dislocations, carbides, etc.) is accounted for in this case study by expanding the reaction term from Eq. (\ref{Eq. R_T trap}) to consider a 3-trap model, 
\begin{equation}
    R_T = -\left[\frac{K_T^c N_T^c/N_L}{(1+K_T^c C_L/N_L)^2}+\frac{K_T^d N_T^d/N_L}{(1+K_T^d C_L/N_L)^2}+\frac{K_T^{gb} N_T^{gb}/N_L}{(1+K_T^{gb} C_L/N_L)^2}\right]\frac{\partial C_L}{\partial t} +\theta_T^d\frac{dN_T^d}{d\varepsilon_p}\frac{\partial \varepsilon_p}{\partial t}
\end{equation}
where the influence of the plastic strain rate in the creation of dislocations is also considered. The predictions obtained are benchmarked against the pioneering results by Dadfarnia et al.  \cite{Dadfarnia2011HydrogenEmbrittlement}. Following Ref. \cite{Dadfarnia2011HydrogenEmbrittlement}, the parameters used are listed in Table \ref{Tab: mat_multitrap}, providing binding energies and trap densities for three types of traps: carbides (superscript $c$), dislocations (superscript $d$), and grain boundaries (superscript $gb$). Trapping constants are calculated from the corresponding binding energy, $K_T^i=\exp(E_B^i/RT)$. The trap density of grain boundaries and carbides remains constant, unlike their dislocations counterpart, which is determined for a bcc microstructure as a function of the total density of dislocations, $\rho^d$ and the lattice parameter $a$:
\begin{equation}
    N_T^d = \frac{\sqrt{2} \rho^d}{a}
\end{equation}
The dislocation density evolution with increasing plastic strain is assumed by Dadfarnia et al. \cite{Dadfarnia2011HydrogenEmbrittlement} from \cite{GilmanJJ.1969MicromechanicsSolids},
\begin{equation}
\label{Eq: rho evolution}
    \rho=
    \begin{cases}
        \rho_0 + 2\gamma\varepsilon_p & \text{if } \varepsilon_p \le 0.5\\
        \rho_0+\gamma & \text{if } \varepsilon_p > 0.5\\
    \end{cases}
\end{equation}
where the dislocation density without plastic deformation, $\rho_0$, equals $10^{10}$ m$^{-2}$ and $\gamma$ is chosen as $10^{16}$ m$^{-2}$ \cite{Dadfarnia2011HydrogenEmbrittlement}.
It must also be noted that the boundary and initial conditions differ from those adopted in the previous Section. Here, a ramp pressure from zero to a $p_{max}=15$ MPa is simulated and the concentration boundary condition follows the corresponding Sievert's law behaviour; thus, the evolution of the boundary condition is given by:
\begin{equation}
C_L(\mathcal{B})=K\sqrt{p_{H_2}}=K\sqrt{\frac{t}{t_{load}}p_{max}}=C_{env}\sqrt{\frac{t}{t_{load}}}
\end{equation}
where $C_{env}$ here substitutes $K\sqrt{p_{max}}$ and its value is given in Table \ref{Tab: mat_multitrap}. Moreover, initial conditions assume an initially empty bulk, i.e. a zero initial hydrogen concentration. This scenario is of interest from a computational perspective as it induces numerical oscillations. 

\begin{table}[H]
\centering
\caption{Parameters employed in the benchmark analysis for the multi-trap hydrogen model, mimicking the work by Dadfarnia et al.  \cite{Dadfarnia2011HydrogenEmbrittlement}.}
\label{Tab: mat_multitrap}
   {\tabulinesep=1.2mm
   \makebox[\textwidth][c]{\begin{tabu} {cccccccccccccc}
       \hline
$E$  & $\nu$ & $\sigma_{y0}$ & $N$ & $T$ & $D_L$  & $\bar{V}_H$ & $C_{env}$  \\ \hline
201.88 & 0.3  & 595 & 0.059 & 300 & 1.27$\times$10$^{-8}$ & 2$\times$10$^{-6}$  & 2.66$\times$10$^{22}$  \\
(GPa) & (-) &  (MPa) & (-) & (K) & (m$^2$/s)  & (m$^3$/mol) & (atoms/m$^3$)    \\ 
        \end{tabu}}}
{\tabulinesep=1.2mm
   \makebox[\textwidth][c]{\begin{tabu} {cccccccccccccc}
   \hline
$N_L$ & $N_T^d$  & $N_T^c$ & $N_T^{gb}$ & $E_B^d$ & $E_B^c$ & $E_B^{gb}$\\ \hline
8.46$\times$10$^{28}$ & $\sqrt{2}\rho/a$ & 10$^{-2}N_L$  & 10$^{-6}N_L$ & 20.2 & 11.5 & 58.6\\
(sites/m$^3$) & (sites/m$^3$)  & (sites/m$^3$)  & (sites/m$^3$) & (kJ/mol) & (kJ/mol) & (kJ/mol)\\ \hline
        \end{tabu}}}
        
\end{table}

\begin{figure}[H]
        \begin{subfigure}[h]{1\textwidth}
                \centering                \includegraphics[scale=0.8]{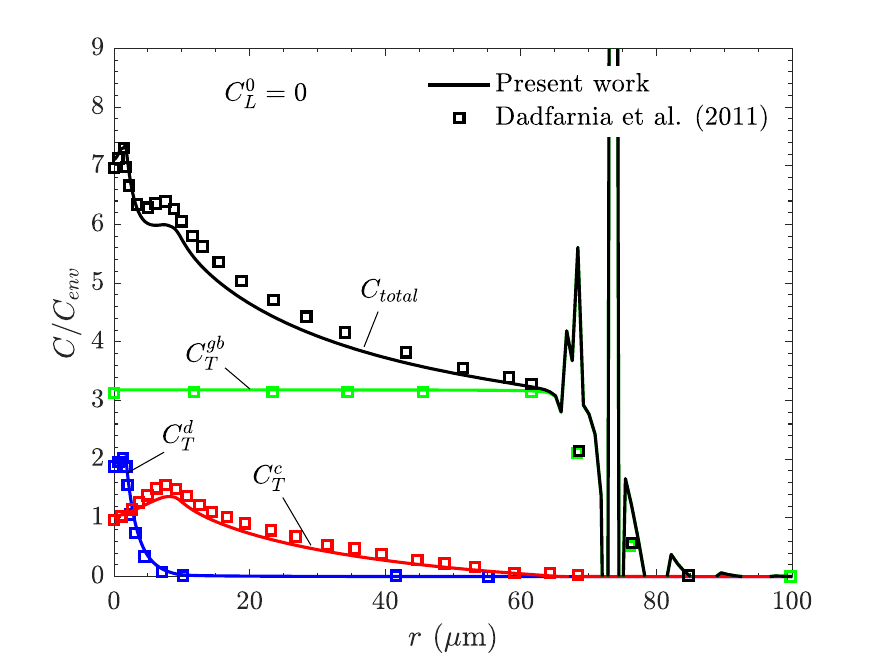}
                \caption{}
                \label{fig: Multitrap_Ctotal_empty}
        \end{subfigure}\\
        \begin{subfigure}[h]{1\textwidth}
                \centering               \includegraphics[scale=0.8]{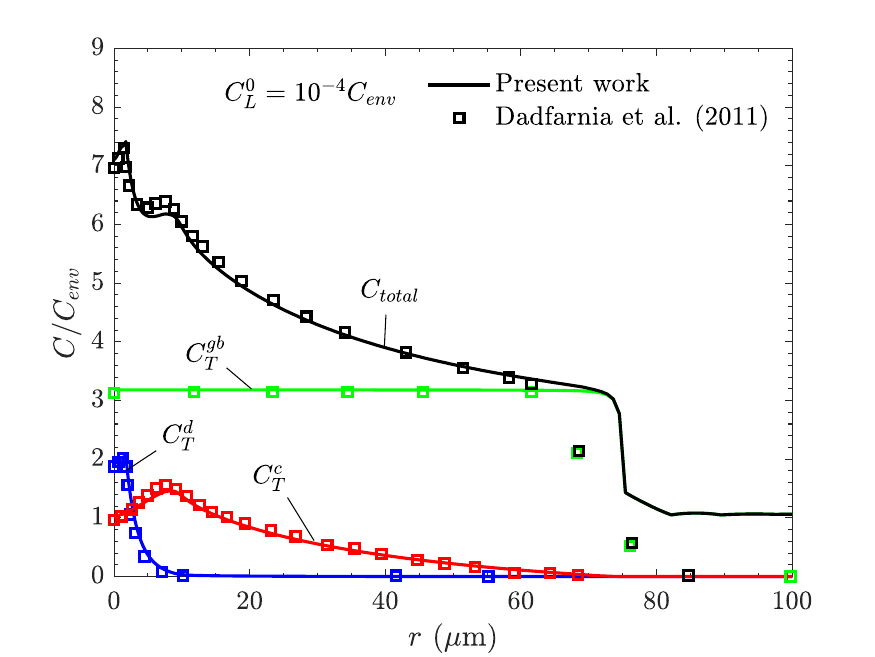}
                \caption{}
                \label{fig: Multitrap_Ctotal_CL01E-4}
        \end{subfigure}       
        \caption{Distributions of total and trapped hydrogen concentrations considering multiple trap types - verification with the results by Dadfarnia et al. \cite{Dadfarnia2011HydrogenEmbrittlement} (symbols). Two scenarios are considered: (a) a hydrogen-free sample at $t=0$, as in Ref. \cite{Dadfarnia2011HydrogenEmbrittlement}, and (b) a sample with a small initial lattice concentration equal to $C_L^0 = 10^{-4} C_{env}$. The agreement is very good near the crack tip but oscillations induce differences far from the crack tip.}
\label{fig: Multitrap_Ctotal_quad}
\end{figure}

Sofronis and McMeeking \cite{Sofronis1989NumericalTip} were the first to point out that starting the analysis with a null hydrogen concentration in the sample led to numerical instabilities and therefore considered only uniformly pre-charged samples in their study. A potential workaround is to define a very small initial concentration ($C_L (t=0) = C_L^0 \approx 0^+$) but, as shown below, results can be sensitive to the magnitude of these small, artificial concentrations, particularly for multi-trap case studies. The predicted hydrogen distribution ahead of the crack tip is given in Fig. \ref{fig: Multitrap_Ctotal_quad}, together with the results by Dadfarnia et al. \cite{Dadfarnia2011HydrogenEmbrittlement} (symbols). The results show the hydrogen concentrations in each of the trap types considered, as well as the total hydrogen concentration, for two scenarios: (i) with a null initial hydrogen concentration in the sample $C_L^0 = 0$ (Fig. \ref{fig: Multitrap_Ctotal_quad}a), and (ii) for a small, residual initial hydrogen content $C_L^0 = 10^{-4} C_{env}$ (Fig. \ref{fig: Multitrap_Ctotal_quad}b). While, in both cases, a good agreement is attained for all quantities near the crack tip, differences and numerical oscillations are observed at distances ahead of the crack tip of \SI{60}{\micro\meter} or larger. The numerical oscillations observed influence the grain boundary concentration (due to the high value of $E_B^{gb}$) and, consequently, the total hydrogen content. 

The influence of the initial hydrogen concentration is more comprehensively investigated in Fig. \ref{fig: Multitrap_CL}, where the lattice hydrogen distribution is shown for various $C_L^0$ choices. The results show that the null condition ($C_L^0 = 0$) results in strong undershoots of negative concentration and this is only solved with artificial values higher than $C_L^0 =10^{-4}C_{env}$. It is also seen that oscillations are locally present, even for initial hydrogen concentrations as high as $C_L^0 =10^{-3}C_{env}$, which do not lead to (unphysical) negative lattice hydrogen contents. Moreover, introducing a \texttt{Lower Limit} does not prevent negative concentrations in some nodes for a quadratic discretization of $C_L$ considered here.

\begin{figure}[H]
  \makebox[\textwidth][c]{\includegraphics[width=0.8\textwidth]{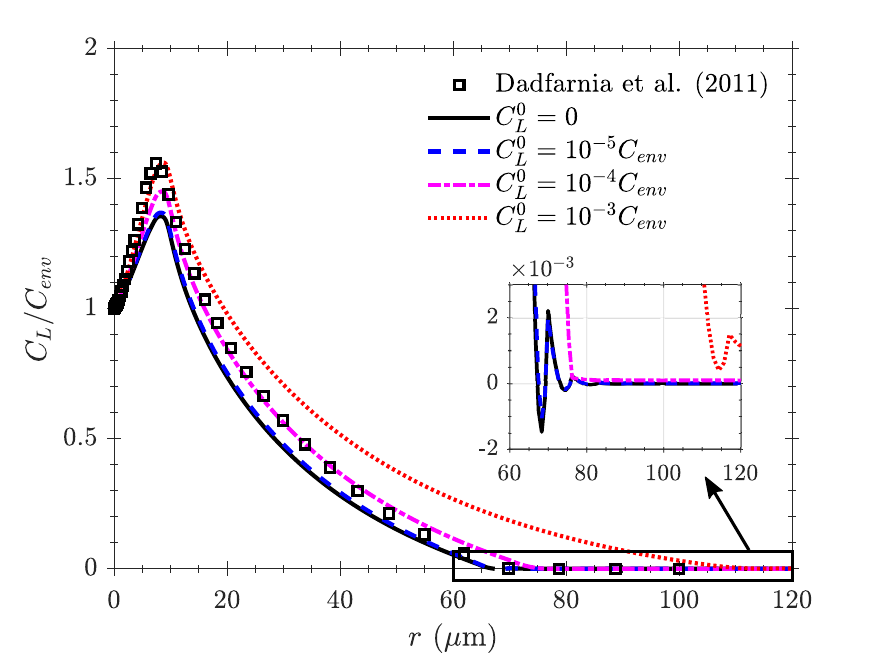}}%
  \caption{Distributions of lattice hydrogen considering a multi-trap scenario \cite{Dadfarnia2011HydrogenEmbrittlement}. A zero initial concentration results in some oscillations but results are sensitive to the choice of an artificial initial hydrogen content.}
  \label{fig: Multitrap_CL}
\end{figure}

\begin{figure}[H]
\makebox[\linewidth][c]{%
        \begin{subfigure}[b]{0.55\textwidth}
                \centering
                \includegraphics[scale=0.65]{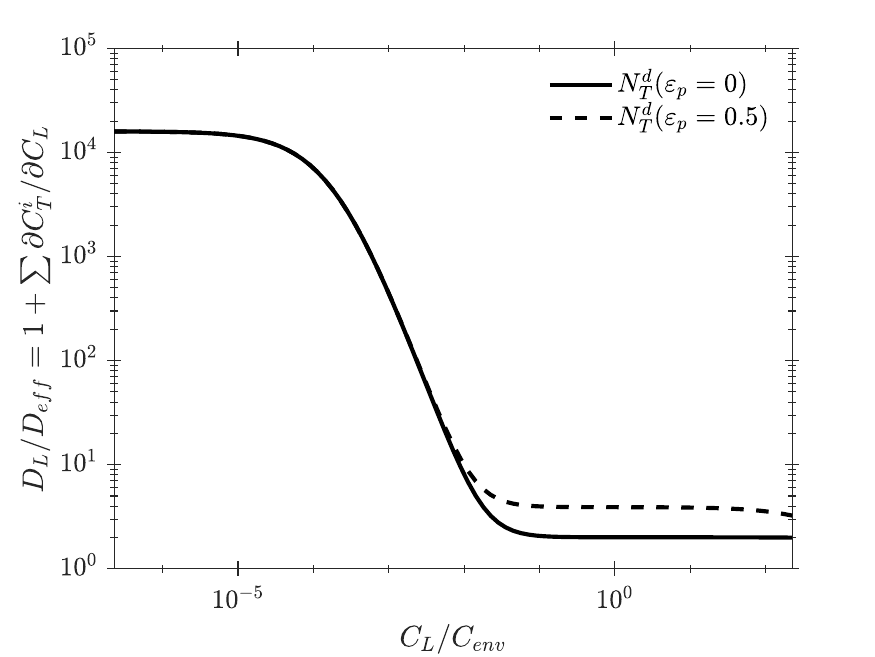}
                \caption{}
                \label{fig:DLDeff}
        \end{subfigure}
        \begin{subfigure}[b]{0.55\textwidth}
                \raggedleft
                \includegraphics[scale=0.65]{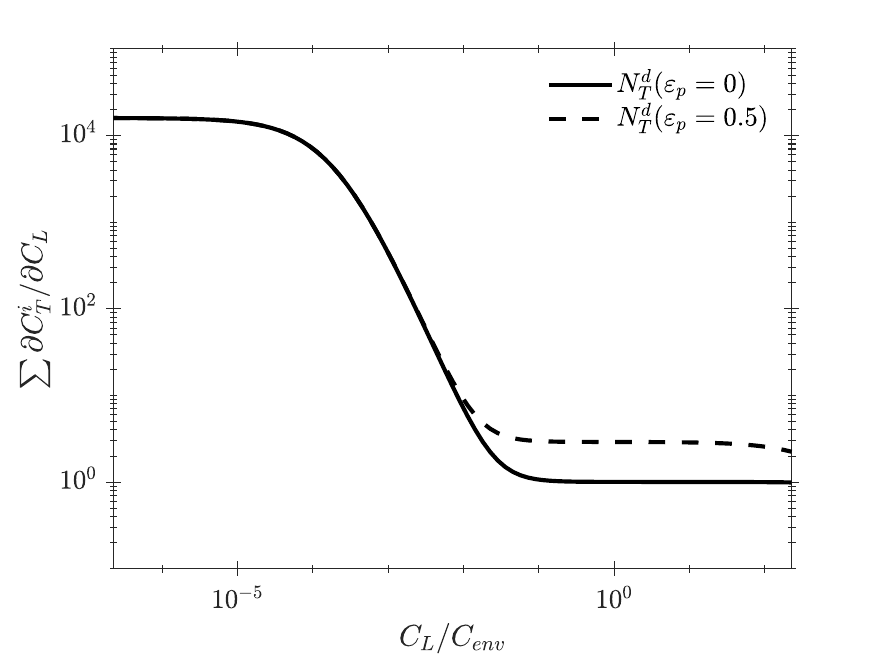}
                \caption{}
                \label{fig:DLDeff_reaction}
        \end{subfigure}}
        \caption{Diffusion delay considering multiple trapping sites, following the work from Dadfarnia et al. \cite{Dadfarnia2011HydrogenEmbrittlement}, showing how the operational diffusivity, $D_L/D_{eff}=1+\sum \partial C_T^i/\partial C_L$, and sink reaction, $\sum \partial C_T^i/\partial C_L$, vary with the normalised lattice hydrogen content ($C_L/C_{env}$).}
        \label{fig:FigDeffDadfarnia}
\end{figure}

Two strategies have been tried to reduce oscillations in the results: (i) consistent stabilization schemes, and (ii) implementing the trapping term as a damping $D_L/D_{eff}$ coefficient in the \texttt{Stabilized Convection-Diffusion Equation} module. However, their role in reducing numerical oscillations is found to be negligible. Streamline and crosswind stabilization schemes do not work for the present case since undershoots are not caused by a convection-diffusion unbalance as it is a reaction-dominated problem \cite{Nadukandi2010AProblem}. Similarly, the reformulation of the reaction term as a damping coefficient, i.e. $D_L/D_{eff}=1+\sum \partial C_T^i/\partial C_L$, does not improve the stability of the solution. It can be thus concluded that numerical oscillations are not due to the steep gradient in $C_L$ but to the reaction-dominated problem and the sensitive character of the trapping term. For strong traps (high $E_B$), the term $\sum \partial C_T^i/\partial C_L$ becomes very high when traps are being filled. For simulations without pre-charging, i.e. $C_L^0 = 0$, as in Dadfarnia et al. \cite{Dadfarnia2011HydrogenEmbrittlement}, there can be a very strong gradient of damping or reaction terms, as shown in Fig. \ref{fig:FigDeffDadfarnia}, where the variation in the terms $D_L/D_{eff}=1+\sum \partial C_T^i/\partial C_L$ and $\sum \partial C_T^i/\partial C_L$ is plotted as a function of the normalised lattice hydrogen content ($C_L/C_{env}$). Both terms change orders over the relevant $C_L$ values. For this reason, numerical noise is significant near the moving front that separates zero and non-zero lattice hydrogen concentrations. Damping and reaction terms are plotted for trap density values corresponding to $\varepsilon_p=0$ and $\varepsilon_p=0.5$, showing a minor influence of plastic deformation. 

Nevertheless, as shown in Fig. \ref{fig: Multitrap_Ctotal_linear}a, the use of linear discretization for the variable $C_L$ significantly reduces numerical oscillations, even for the $C_L^0=0$ case. Furthermore, as shown in Fig. \ref{fig: Multitrap_Ctotal_linear}b, the imposition of a \texttt{Lower Limit} constraint avoids negative concentrations in all nodes, in contrast to what was observed when using a quadratic discretization for $C_L$. When comparing Figs. \ref{fig: Multitrap_Ctotal_linear}a and \ref{fig: Multitrap_Ctotal_linear}b, one can see that the total hydrogen concentration is slightly higher for the latter case, showing that negative concentrations do not only influence local results near the $C_L$ front but have a wider influence, acting as a barrier for hydrogen transport.

\begin{figure}[H]
        \begin{subfigure}[h]{1\textwidth}
                \centering                \includegraphics[scale=0.8]{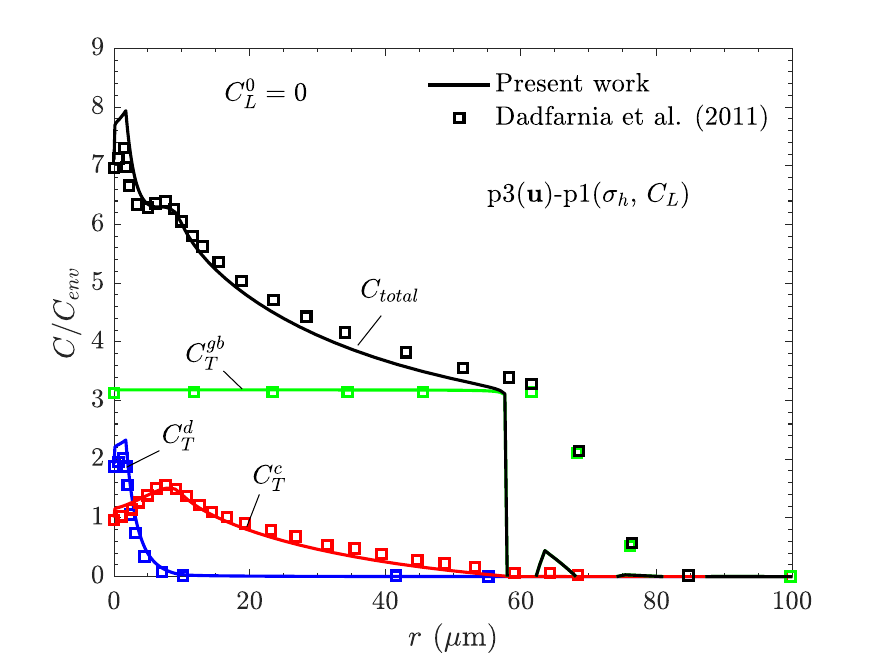}
                \caption{}
                \label{fig: Multitrap_Ctotal_linear_empty}
        \end{subfigure}\\
        \begin{subfigure}[h]{1\textwidth}
                \centering               \includegraphics[scale=0.8]{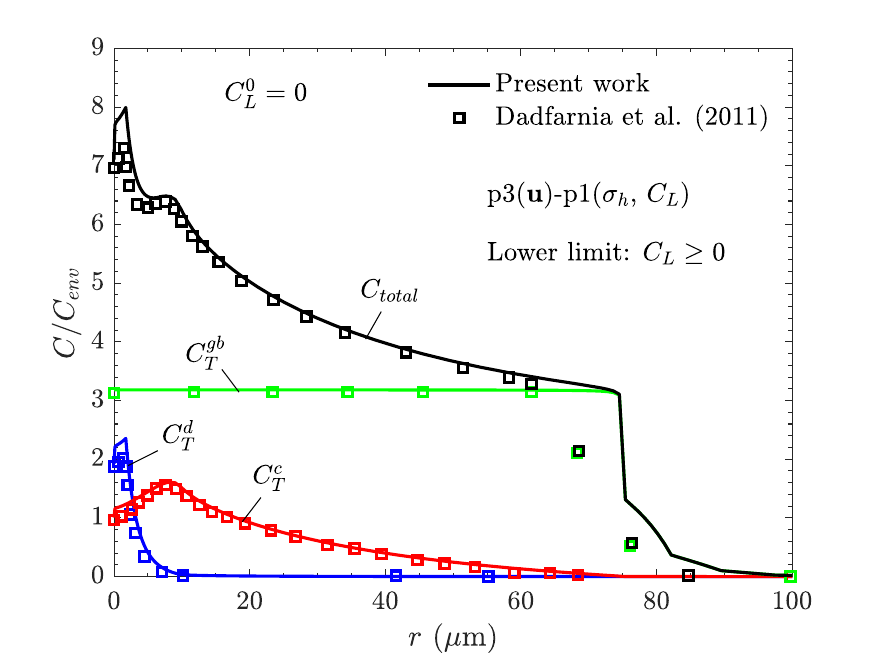}
                \caption{}
                \label{fig: Multitrap_Ctotal_linear_empty_LowerLimit}
        \end{subfigure}       
        \caption{Distributions of total and trapped hydrogen concentrations considering multiple trap types - verification with the results by Dadfarnia et al. \cite{Dadfarnia2011HydrogenEmbrittlement} (symbols). The results in (a) show that the use of a linear (p1) discretization for $C_L$ significantly reduces oscillations (relative to Fig. \ref{fig: Multitrap_Ctotal_quad}a), while the results in (b), show that establishing a \texttt{Lower Limit} constraint on $C_L$ is effective with this discretisation, bringing the results much closer to those by Dadfarnia et al. \cite{Dadfarnia2011HydrogenEmbrittlement}.}
\label{fig: Multitrap_Ctotal_linear}
\end{figure}

\subsection{Case 3: Hydrogen transport by dislocations}
\label{Sec:Hdislocations}

Here, we consider the assumption that hydrogen atoms can be enhanced via dislocation mobility, as described in Section \ref{Sec:TheoryDislocationsH} and first considered by Dadfarnia et al. \cite{Dadfarnia2014ModelingDislocations}. To validate the implementation through a convective term, the dislocation motion direction is assumed to be parallel to the crack plane, from the tip to the bulk; i.e., the vector $ \textbf{n}= (1,0)$ is considered for the flux term described in Eq. (\ref{Eq. dislocation flux}). In addition, two acceleration factors $n_d$ due to hydrogen are reproduced. The parameters characterising the elastoplastic and diffusion material behaviour are taken from Ref. \cite{Dadfarnia2014ModelingDislocations} and listed in Table \ref{Tab:mat_Dadafarnia_dislocations}. Mimicking Ref. \cite{Dadfarnia2014ModelingDislocations}, and in contrast to the previous case studies, the material properties adopted correspond to those typical of X70 or X80 pipeline steel.

\begin{table}[H]
\centering
\caption{Parameters used for the validation of hydrogen transport by dislocations, based on the work by Dadfarnia et al. \cite{Dadfarnia2014ModelingDislocations}.}
\label{Tab:mat_Dadafarnia_dislocations}
   {\tabulinesep=1.2mm
   \makebox[\textwidth][c]{\begin{tabu} {cccccccccccccc}
       \hline
$E$  & $\nu$ & $\sigma_{y0}$ & $N$ & $T$ &   \\ \hline
200 & 0.3  & 600 & 0.06 & 300  \\
(GPa) & (-) &  (MPa) & (-) & (K)    \\ \hline
$D_L$  & $\bar{V}_H$ & $N_L$ & $C_{env}, C_L^0$ & $E_B$\\ \hline
2$\times$10$^{-8}$ & 2$\times$10$^{-6}$  & 8.46$\times$10$^{28}$ & 2.08$\times$10$^{21}$ & 50 \\
(m$^2$/s)  & (m$^3$/mol) & (sites/m$^3$) & (atoms/m$^3$) & kJ/mol\\ \hline
        \end{tabu}}}
\end{table}

The lattice parameter $a$ equals 0.287 nm and the Burgers vector $b_v = $ 0.248 nm \cite{Dadfarnia2014ModelingDislocations}. For the trap density evolution as a function of plastic strain, the expression from Kumncick and Johnson  \cite{Kumnick1980DeepIron} was used, Eq. (\ref{Eq.Kumnick}). It is also assumed, as in Ref. \cite{Dadfarnia2014ModelingDislocations}, that dislocation traps can accommodate 10 hydrogen atoms; i.e. the trap density is $\alpha N_T$ with $\alpha = 10$ and $N_T$ being given by Eq. (\ref{Eq.Kumnick}). In this benchmark, a ramp load of 0.01 MPa$\sqrt{\text{m}}$/s was simulated, until a maximum value of $K_I$ = 100 MPa$\sqrt{\text{m}}$ was reached.\\ 

\begin{figure}[H]
  \makebox[\textwidth][c]{\includegraphics[width=0.8\textwidth]{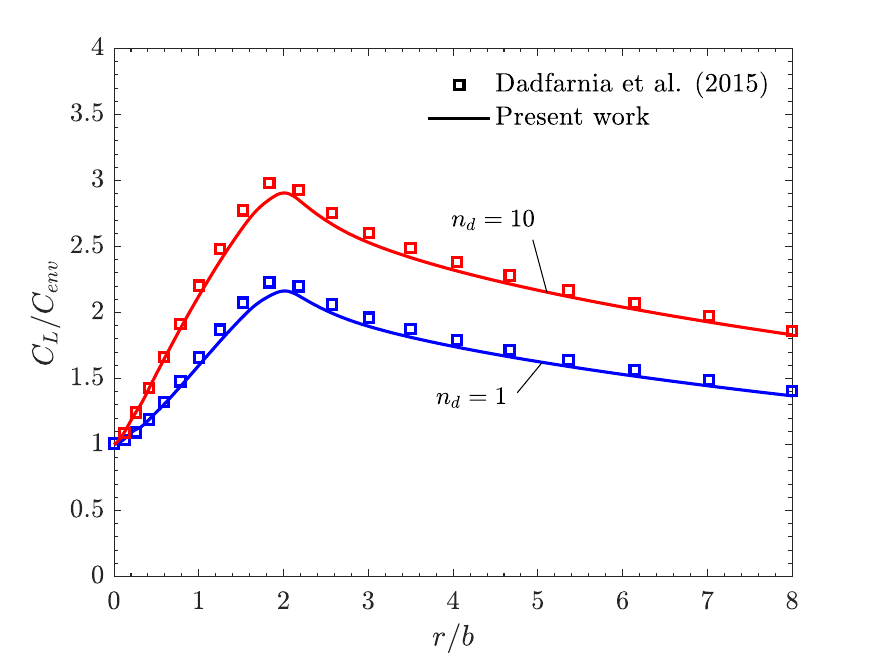}}%
  \caption{Validation of hydrogen transport by dislocations. Comparison of the lattice hydrogen distribution ahead of the crack tip predicted by the present framework and that estimated by Dadfarnia et al.  \cite{Dadfarnia2014ModelingDislocations}. A good agreement is attained, also when considering for the role of the factor $n_d$, which is aimed at capturing a potential enhancement on dislocation velocity.}
  \label{fig:Fig_Dadfarnia_disl}
\end{figure}

The results obtained are shown in Fig. \ref{fig:Fig_Dadfarnia_disl}. The normalised lattice hydrogen concentration is plotted as a function of the normalised distance ahead of the crack tip, as predicted by both the present framework and the work by Dadfarnia et al. \cite{Dadfarnia2014ModelingDislocations}. A decent agreement is observed, with the convective term capturing hydrogen transport by dislocations, as well as the increase in the distribution of hydrogen concentration in lattice sites when the dislocation velocity is multiplied by $n_d$.

\subsection{Case 4: Kinetic trapping}
\label{Sec:MFresults}

We proceed to validate the ability of the framework to predict kinetic trapping, as per McNabb and Foster's model. As detailed in Section \ref{Sec: Physics PDE McNabb}, an additional PDE, Eq. (\ref{eq. MCNabb-Foster}), is implemented to capture kinetic trapping. In this case, a quadratic discretization is chosen for the additional dependent variable ($C_T$ or $\theta_T$). As presented in Section \ref{Sec: Physics PDE McNabb}, two approaches can be followed to model trapping kinetics, implementing McNabb and Foster's reaction as: (i) a function trap occupancy rate $\theta_T$ (Option 1), or (ii) as a function of hydrogen concentration in trapping sites $C_T$ (Option 2). The first benchmark, based on the work by Mart\'{\i}nez-Pa\~neda al. \cite{Martinez-Paneda2020GeneralisedTips}, aims at validating Option 1, while the second benchmark, based on the work by Charles et al. \cite{Charles2021EffectTip}, aims at validating Option 2.\\

We begin by describing the validation of our implementation of Option 1 against the results by Mart\'{\i}nez-Pa\~neda et al. \cite{Martinez-Paneda2020GeneralisedTips}, which were obtained using a constant concentration as a boundary condition and the same parameters as the classic benchmark by Sofronis and McMeeking \cite{Sofronis1989NumericalTip} reproduced in Section \ref{Sec. Sofronis and Krom benchmark}. That is, the parameters listed in Table \ref{Tab:mat_Sofronis}, but including McNabb \& Foster's formulation for kinetic trapping. The trapping constant, here called $\kappa$ in Eq. (\ref{eq. MCNabb-Foster}), was taken to be $1.68\times 10^8$ s$^{-1}$, which is equivalent to $k_r=\kappa/N_L=3.3\times 10^{22}$ m$^{3}$s$^{-1}$sites$^{-1}$ in the reference paper by Mart\'{\i}nez-Pa\~neda et al. \cite{Martinez-Paneda2020GeneralisedTips}. Thus, the detrapping constant $\lambda$ is determined by the trapping constant and the magnitude of the binding energy through the equilibrium constant: $\lambda = \kappa/K_T$.\\

\begin{figure}[H]
        \begin{subfigure}[h]{1\textwidth}
                \centering                \includegraphics[scale=0.8]{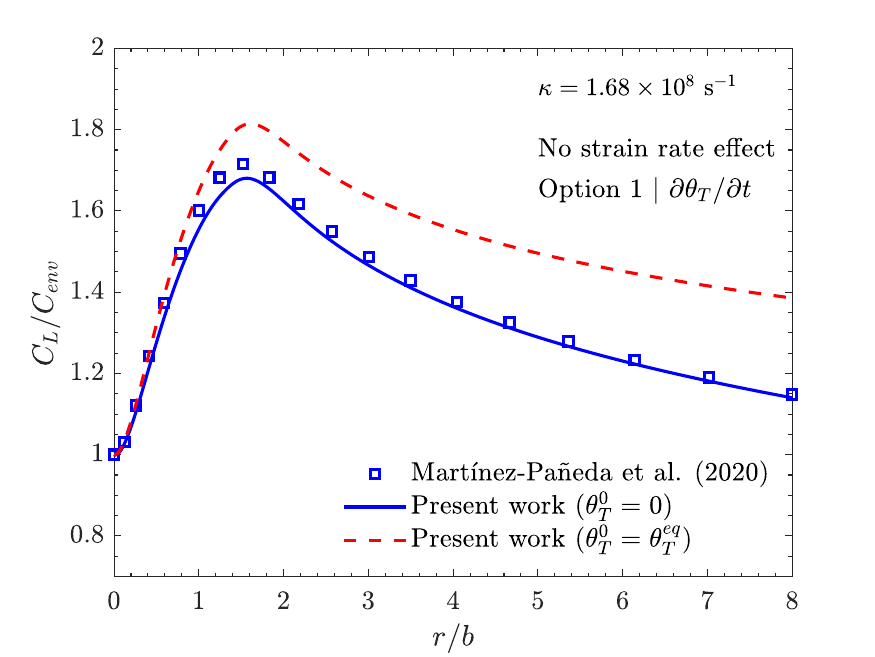}
                \caption{}
                \label{fig: CL_McNabb_Paneda2020_a_Validation}
        \end{subfigure}\\
        \begin{subfigure}[h]{1\textwidth}
                \centering               \includegraphics[scale=0.8]{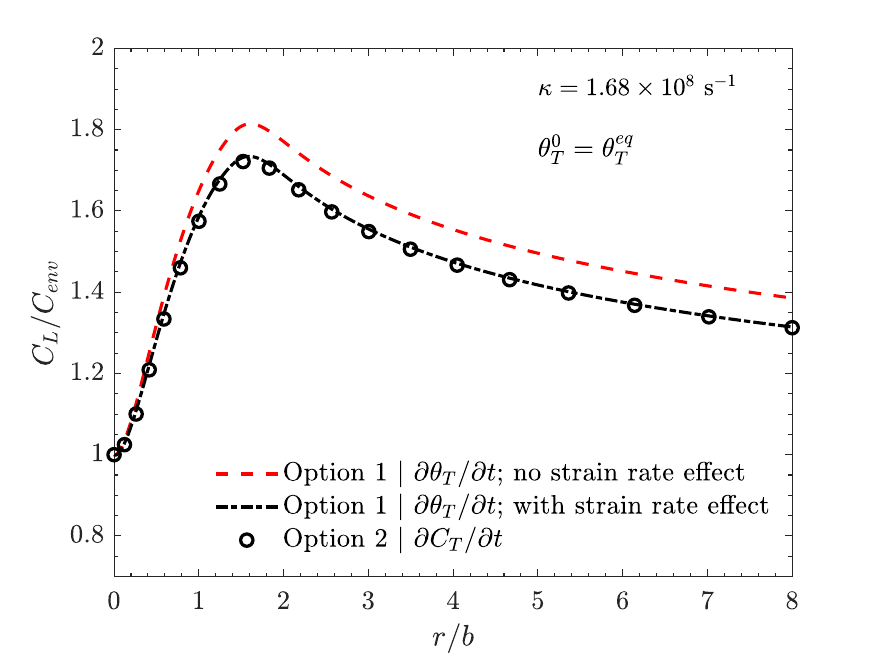}
                \caption{}
                \label{fig: CL_McNabb_Paneda2020_b_Strain_rate}
        \end{subfigure}       
        \caption{Modelling kinetic trapping (McNabb and Foster formulation): (a) validation against the results by Mart\'{\i}nez-Pa\~neda et al. \cite{Martinez-Paneda2020GeneralisedTips} under the assumption of traps being empty at $t=0$ and consideration of the role of an initial trap occupancy, as described by equilibrium; and (b) demonstration that solving trapping kinetics using $C_T$ as primary variable implicitly accounts for the interplay between strain rates and trap creation kinetics first highlighted by Krom et al. \cite{Krom1999HydrogenTip}.}
\label{fig:Fig_McNabb_Paneda}
\end{figure}

In Fig. \ref{fig: CL_McNabb_Paneda2020_a_Validation}, the implementation of Option 1 is validated against the work by Mart\'{\i}nez-Pa\~neda et al. \cite{Martinez-Paneda2020GeneralisedTips} under the assumption of traps being initially empty ($\theta_T^0 = 0$). A good agreement is observed. Next, we proceed to assess the influence of the choice of initial trapping conditions as results are sensitive to the choice of $\theta_T^0$ (or $C_T^0$), in contrast with the Oriani-based approach where equilibrium is used to determine the initial trap occupancy from $C_L^0$. For the present benchmark problem, where $C_L^0>0$, the $\theta_T^0=0$ condition is not physically representative, since a uniform pre-charging will produce near-equilibrium trap concentrations. This is also illustrated in Fig. \ref{fig: CL_McNabb_Paneda2020_a_Validation}, where the result obtained when $\theta_T^0$ is determined from Oriani's equilibrium is presented with a dashed red curve. It can be seen that the $C_L$ peak increases with the resulting increased trap occupancy, as lattice hydrogen is otherwise needed to kinetically fill traps in the $\theta_T^0 = 0$ scenario. Another consideration is that the results shown in Fig. \ref{fig: CL_McNabb_Paneda2020_a_Validation} do not consider the so-called Krom term; i.e., how the trap creation rate evolves with plastic strain rate - see Eq. (\ref{eq:KromTermImpl}). We show in Fig. \ref{fig: CL_McNabb_Paneda2020_b_Strain_rate} that the results obtained with Options 1 and 2 are identical when Option 1 (solving for $\partial \theta_T / \partial t$) incorporates Krom's term. In other words, when solving trapping kinetics using $\partial C_T / \partial t$ (Option 2), the trap creation rate term is implicitly accounted for, but it must be included in the reaction term when McNabb and Foster's equation is expressed as $\partial \theta_T / \partial t$ (Option 1 with strain rate effect):

\begin{equation}
    R_T = -N_T\frac{\partial \theta_T}{\partial t} - \theta_T\frac{dN_T}{d\varepsilon_p}\frac{\partial \varepsilon_p}{\partial t}
\end{equation}

In addition, the results shown in Fig. \ref{fig: CL_McNabb_Paneda2020_b_Strain_rate} reveal that removing Krom's strain rate effect results in an increase in the lattice hydrogen concentration, consistent with the results obtained in the Oriani-based analysis (see Fig. \ref{fig:Fig_SM_Krom}).

\begin{figure}[H]
  \makebox[\textwidth][c]{\includegraphics[width=0.8\textwidth]{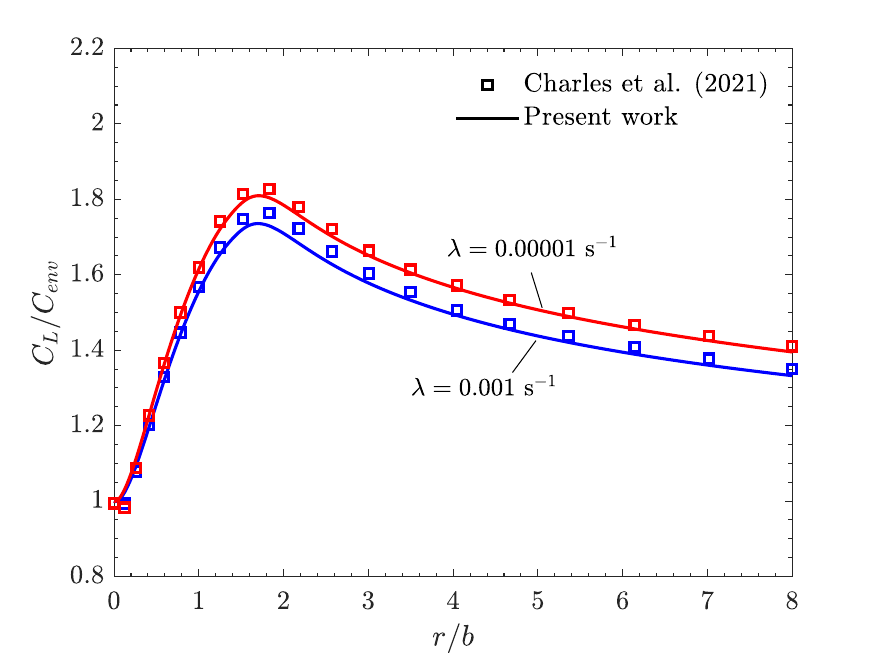}}%
  \caption{Modelling kinetic trapping (McNabb and Foster formulation): validation against the results by Charles et al. \cite{Charles2021EffectTip}, implementing kinetic trapping through the term $R=-\partial C_T / \partial t$ (Option 2), as discussed in Section \ref{SEc:Option2_Charles}.}
  \label{fig:Fig_Charles}
\end{figure}

The second case study regarding kinetic trapping reproduces the work by Charles et al. \cite{Charles2021EffectTip}, so as to verify the approach of implementing McNabb and Foster's formulation only through the term $R = -\partial C_T / \partial t$, and eliminating the explicit consideration of the strain rate term (i.e., so-called Option 2 in Section \ref{Sec: Physics PDE McNabb}). Here, we obtain crack tip $C_L$ distributions for two different detrapping constants $\lambda$, with the trapping constant $\kappa$ being estimated from the choice of $\lambda$ and the binding energy. The remaining parameters follow the classic studies by Sofronis and McMeeking \cite{Sofronis1989NumericalTip} and Krom et al. \cite{Krom1999HydrogenTip}. The results, shown in Fig. \ref{fig:Fig_Charles}, reveal an excellent agreement between the present implementation and the work by Charles et al. \cite{Charles2021EffectTip}. It must be noted that in their work an analytical approximation of McNabb and Foster's equation was used to solve $\theta_T$, eliminating the need for an additional degree of freedom, following the implementation proposed by Benannoune et al. \cite{Benannoune2018NumericalEquation}. 
 
\subsection{Case 5: Chemical potential and suitable boundary conditions}
\label{Sec:ChemPotResults}

Previous examples reproduced works where a constant concentration, $C_L=C_{env}$, was imposed as a boundary condition on the crack surface. However, it has already been shown that this is not consistent with the thermodynamics of hydrogen uptake from a gaseous hydrogen source - the hydrostatic stress-dependency of the solubility must be accounted for. Two approaches can be followed to account for this more rigorous description of the surface conditions: defining a stress-dependent concentration boundary condition, as per Eq. (\ref{Eq:CL_stress}), or using the chemical potential $\mu_L$ as the primal variable, as described in Sections \ref{Sec. Stress-dependent BCs} and \ref{mu_L implementation}. Both approaches are implemented here, compared and validated against the $\mu_L$-based implementation by Di Leo and Anand \cite{DiLeo2013HydrogenDeformations}. This boundary value problem mimics the analysis by Krom et al. \cite{Krom1999HydrogenTip} and thus all the material parameters, including the values for $C_L^0$ and $C_{env}$, are those listed in Table \ref{Tab:mat_Sofronis} and used in Section \ref{Sec. Sofronis and Krom benchmark} (Case study 1).\\

\begin{figure}[H]
  \makebox[\textwidth][c]{\includegraphics[width=0.8\textwidth]{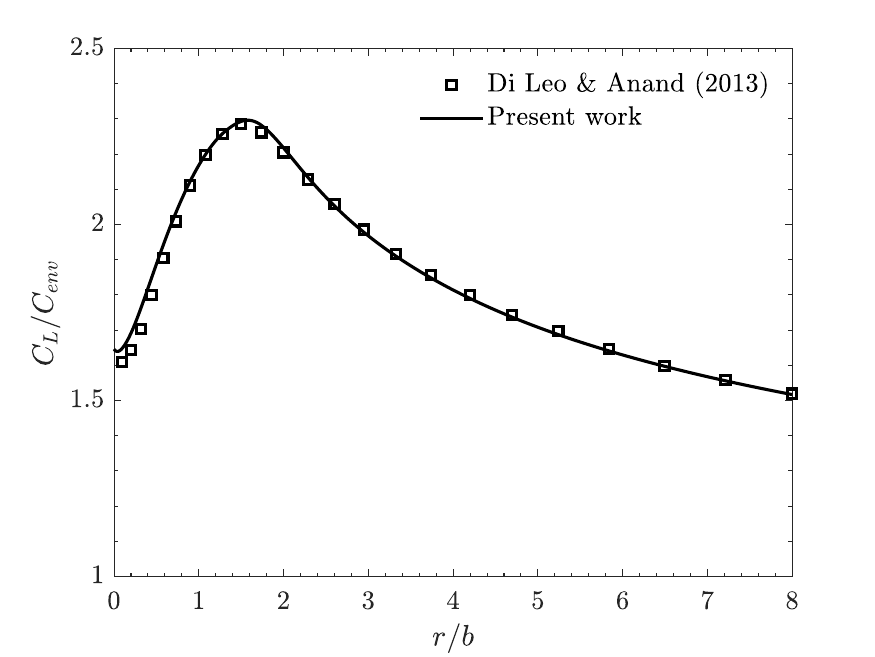}}%
  \caption{Accounting for the role of hydrostatic stresses in augmenting hydrogen uptake. Calculations of lattice hydrogen crack tip distribution obtained with a concentration-based formulation, validating the implementation against 
 the chemical potential-based model by Di Leo and Anand \cite{DiLeo2013HydrogenDeformations}.}
  \label{fig:Fig_DiLeo}
\end{figure}

First, in Fig. \ref{fig:Fig_DiLeo}, the $C_L$-based implementation is validated against the crack tip hydrogen distributions by Di Leo and Anand \cite{DiLeo2013HydrogenDeformations}. A very good agreement is obtained and, as expected, the lattice hydrogen concentration at the metal surface ($r=0$) exceeds $C_{env}$. The results show that the concentration-based approach is a valid alternative to the chemical-potential formulation if enriched with the appropriate boundary condition. It must be mentioned that Dirichlet boundary conditions for diffusion in COMSOL Multiphysics can be applied as an elemental or nodal constraint. In both cases, constraints are added to the surface nodes, but choosing an element approach avoids problems related to intersecting surfaces with different conditions. Although not shown here for the sake of brevity, we have conducted numerical tests and observed that both boundary condition strategies yield the same results.

\begin{figure}[H]
  \makebox[\textwidth][c]{\includegraphics[width=0.8\textwidth]{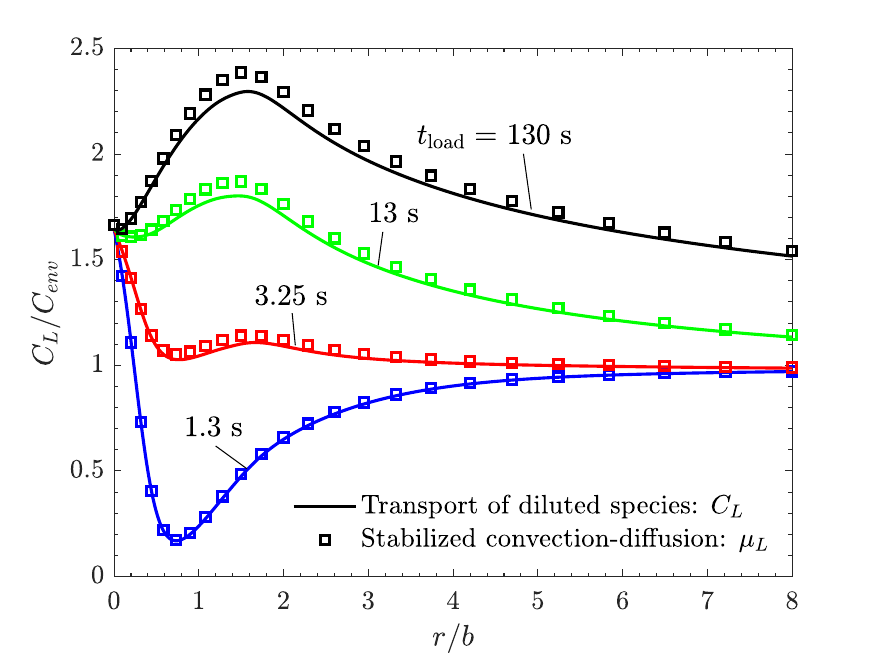}}%
  \caption{Comparing $C_L$- and $\mu_L$-based implementations, with the former having a $\sigma_H$-dependent boundary condition. The hydrogen crack tip distribution results, obtained for different loading times, show a good agreement overall.}
  \label{fig:Fig_muL}
\end{figure}

Next, we proceed to compare the $C_L$-based implementation, which uses the \texttt{Transport of diluted species} interface, with our own $\mu_L$-based implementation, which uses the \texttt{Stabilized convection-diffusion} interface and circumvents the need for stress-dependent boundary conditions. The results are shown in Fig. \ref{fig:Fig_muL}, revealing a very good agreement. Slight deviations are observed in the slowest numerical tests, where the influence of the $\sigma_h$ term is expected to be most significant.

\begin{figure}[H]
  \makebox[\textwidth][c]{\includegraphics[width=0.8\textwidth]{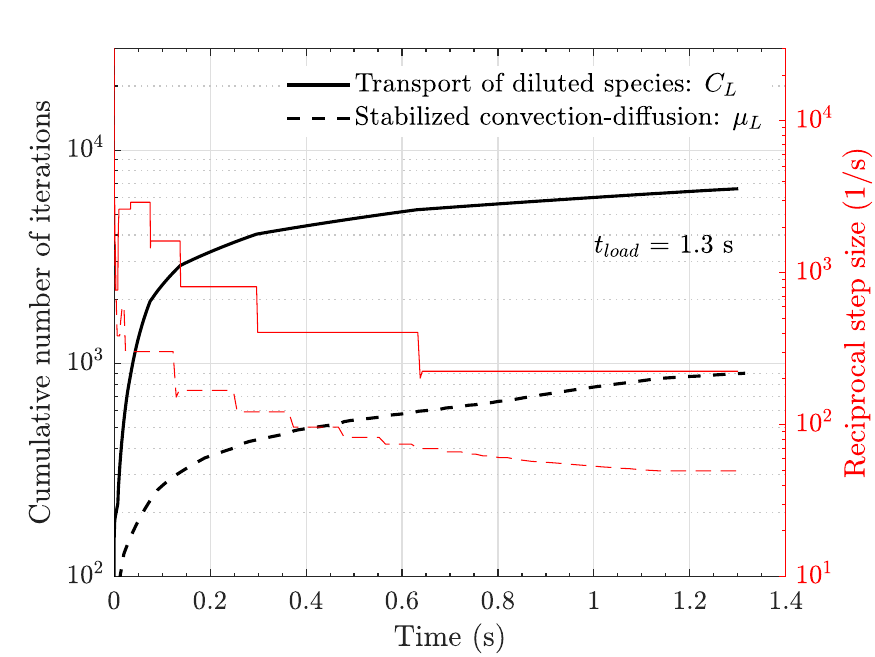}}%
  \caption{Convergence behaviour of the implementations using $C_L$ or $\mu_L$ as the solution variable: cumulative number of iterations (black curves) and inverse (reciprocal) of the step increment (red curves). The $\mu_L$-based implementation shows a smaller number of iterations and larger solver step increments.}
  \label{fig:Fig_muL:convergence}
\end{figure}

Differences in convergence and efficiency are assessed in Fig. \ref{fig:Fig_muL:convergence}. The black curves show the cumulative number of iterations, while the red curves denote the inverse (reciprocal) of the step size (i.e., lower values indicating larger steps and thus better convergence). It can be seen that the $\mu_L$-based implementation is more efficient and robust, displaying faster convergence. The total number of increments is equal to 731 and 150 for the $C_L$ and $\mu_L$-based implementations, respectively. It must be noted that a \texttt{Free} time stepping is selected and therefore the increment size is increased progressively by the solver. Both schemes show a robust convergence with a stable increase in the cumulative number of iterations. Computation times equal 1.23$\times10^4$ and 2.06$\times10^3$ s for the $C_L$- and $\mu_L$-based implementations, respectively. Interestingly, similar computation times are obtained when $\sigma_h$ is treated as an external variable for the $\mu_L$-based case, suggesting that differences in computation times are related to the storage of $\sigma_h$, which appears to influence convergence, and not to the convergence of the transport equations. The $\mu_L$-based strategy circumvents the hydrostatic stress mapping and gradient calculation, alleviating computation costs. It is also worth noting that the scaling of the solution vector has a high influence on convergence; for the purpose of establishing a fair comparison, both $C_L$ and $\mu_L$ dependent variables are scaled using their initial values. These results are shown only for the case of $t_{load}$ = 1.3 s, but a similar outcome is obtained for slower deformation rates.

\begin{figure}[H]
  \makebox[\textwidth][c]{\includegraphics[width=0.8\textwidth]{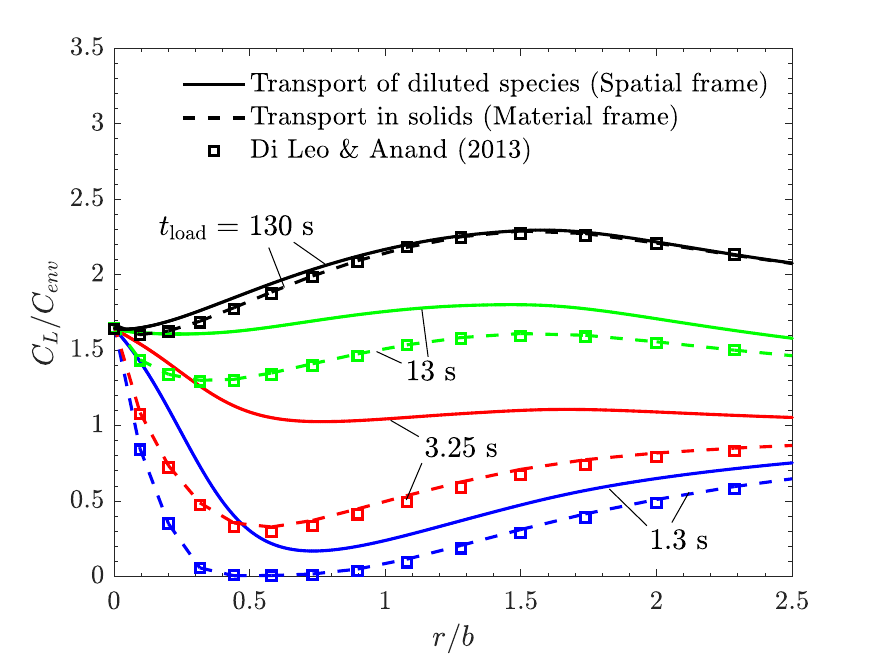}}%
  \caption{On the use of the material or spatial frame to solve the transport equations: crack tip lattice hydrogen distribution for different loading rates, showing that only the consideration of the material frame (as in-built in the \texttt{Transport in solids} interface) delivers a good agreement with the results by Di Leo and Anand \cite{DiLeo2013HydrogenDeformations}.}
  \label{fig:Material_Frame}
\end{figure}

It is important to note that the diffusion equation, including stress-driven diffusion and plastic-strain modified trapping, is solved in a spatial frame when the \texttt{Transport of diluted species} or the \texttt{Stabilized convection-diffusion equation} modules are used. Spatial gradients have also been used for hydrogen transport in previous works exploiting user-defined heat transfer \cite{Diaz2016CoupledAnalogy,del2017cohesive, Diaz2024ExplicitMetals}. However, diffusion coupled to large deformations has been typically solved in a material reference frame by different authors in the framework of continuum thermodynamics \cite{Chester2015AGels, DiLeo2013HydrogenDeformations}, while other authors have considered fluxes in the spatial frame \cite{Duda2010ASwelling, Hong2008AGels}. The use of the material or the spatial frame to solve the mass balance equations can lead to differences when the loading rate is high \cite{Diaz2024ExplicitMetals}. This is shown in Fig. \ref{fig:Material_Frame}, where it can be seen that when the load ($K_I$ = 89.2 MPa$\sqrt{\text{m}}$) is applied over a short time, a good agreement with the results by Di Leo and Anand \cite{DiLeo2013HydrogenDeformations} can only be attained when using the \texttt{Transport in Solids} module, which solves the transport equation in the material frame.

Finally, we consider the importance of an appropriate description of the environment-material interface by considering the conditions of exposure to an aqueous electrolyte. As discussed in Section \ref{Sec:GeneralisedBCs}, the hydrostatic stress also plays a role during electrolytic charging since the absorption constant, $k_{abs}$, is multiplied by a stress-dependent term to account for the enhanced solubility, as modelled in Eq. (\ref{Eq: Absorption}). This more rigorous description of the hydrogen evolution and surface reactions is implemented by means of a Neumann boundary condition, typically referred to as a generalised flux. That is, Eq. (\ref{Eq: Absorption}) is adopted to prescribe the scalar value $J$ for the lattice flux, i.e. $\textbf{J}_L\cdot\textbf{n}=J$, along the crack surfaces. The outcome of the simulations is validated against the work by Mart\'{\i}nez-Pa\~neda et al. \cite{Martinez-Paneda2020GeneralisedTips}. The reaction rate constants characterising the absorption/desorption and hydrogen evolution reactions, Eqs. (\ref{Eq: Absorption}) and (\ref{Eq: Adsorption}), are taken from Refs. \cite{Martinez-Paneda2020GeneralisedTips,Turnbull1996ModellingTip} and listed in Table \ref{Tab:generalised BCs constants}. As in previous examples, the remaining mechanical and hydrogen-related parameters mimic the work by Sofronis and McMeeking \cite{Sofronis1989NumericalTip}. 

\begin{table}[H]
\centering
\caption{Absorption, desorption, charging and recombination constants for the crack wall and tip, following Refs. \cite{Martinez-Paneda2020GeneralisedTips, Turnbull1996ModellingTip}.}
\label{Tab:generalised BCs constants}
   {\tabulinesep=1.2mm
   \makebox[\textwidth][c]{\begin{tabu} {ccccc}
       \hline
& $k_{abs}^*$  & $k_{des}$ & $k_c$  & $k_{r,chem}$ \\
& [m/s] & [m/s]  & [mol/(m$^2$s)] &  [mol/(m$^2$s)]\\ 
\hline
Crack wall & $1.18\times10^5$ & $8.8\times10^9$ & $5\times10^{-7}$ & 22\\
Crack tip & $1.18\times10^5$ & $8.9\times10^9$ & $5\times10^{-6}$ & 22\\\hline
   \end{tabu}}}
\end{table}

\begin{figure}[H]
  \makebox[\textwidth][c]{\includegraphics[width=0.8\textwidth]{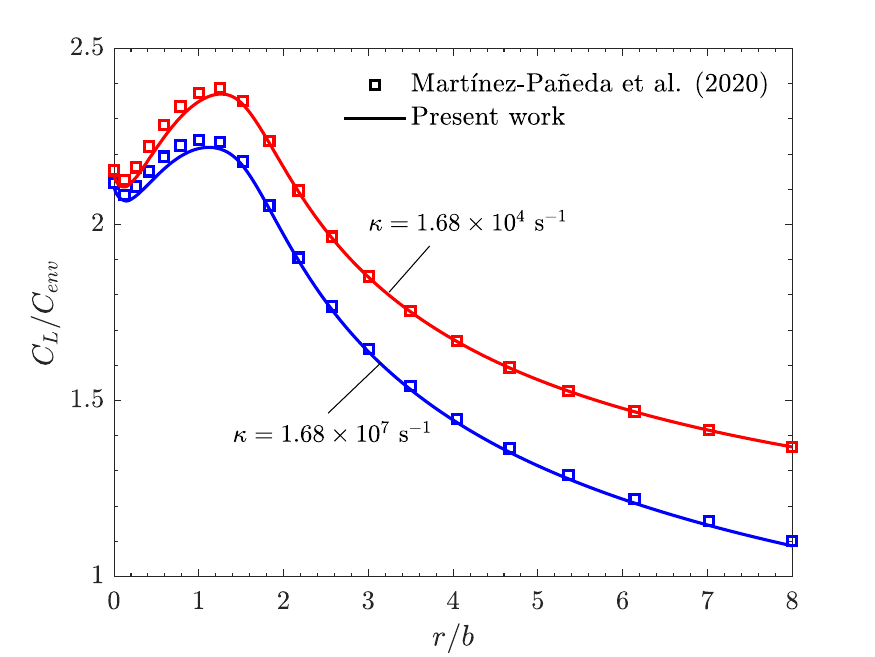}}%
  \caption{Generalised boundary conditions to resolve electrochemical hydrogen uptake: Distributions of interstitial hydrogen for two values of the trapping constant $\kappa$, validating the implementation with Ref. \cite{Martinez-Paneda2020GeneralisedTips}.}
  \label{fig:FigGF}
\end{figure}

The results obtained are shown in Fig. \ref{fig:FigGF}, considering two values of the trapping constant $\kappa$. A very good agreement is obtained. It is also worth noting that in this analysis, the surface magnitude of $C_L$ is not only determined by the stress-dependency of the absorption reaction but also by the constants governing the uptake fluxes. 

\subsection{Case 6: Hydrogen-induced softening}
\label{Sec:Hsoftening}

\begin{figure}[H]
  \makebox[\textwidth][c]{\includegraphics[width=0.8\textwidth]{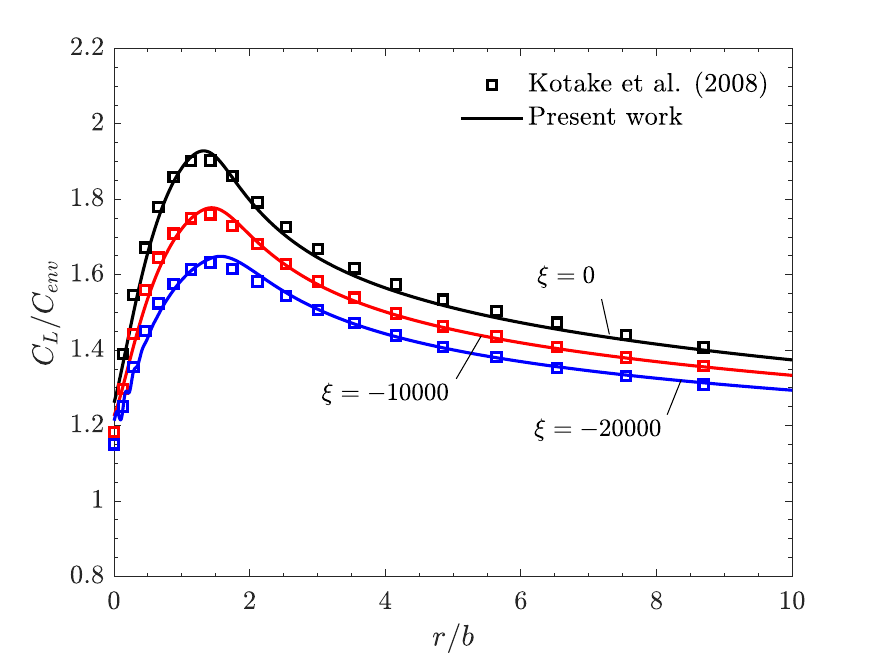}}%
  \caption{Crack tip lattice hydrogen distributions predicted considered hydrogen-induced softening: results from the present framework and validation against the results by Kotake et al. \cite{Kotake2008TransientLoading} for various choices of the hydrogen softening coefficient $\xi$.}
  \label{fig: Kotake_softening}
\end{figure}

The final case study showcases the ability of our generalised framework to capture hydrogen-induced softening. Kotake et al. \cite{Kotake2008TransientLoading} studied the influence of hydrogen-induced softening on the transport of hydrogen near a crack tip in a sample that is being cyclically loaded, and this work is here used as a benchmark to validate our fully coupled implementation. The material parameters are the same as in Section \ref{Sec. Sofronis and Krom benchmark} (Table \ref{Tab:mat_Sofronis}) but the loading conditions differ - the load is increased linearly up to $K_I$ = 40 MPa$\sqrt{\text{m}}$, which is reached after 100 s. It should also be noted that Kotake et al. \cite{Kotake2008TransientLoading} considered a softening coefficient $\xi$ that is related to our softening coefficient $\zeta$, see Eq. (\ref{Eq. H-induced softening}), by $\xi = 1-\zeta$. Thus, a negative value of $\xi$ denotes hydrogen-induced softening. In addition, and in contrast to the previous case studies, in this example the crack surface is assumed to be insulated, which is modelled by the equation:
\begin{equation}\label{Eq:Flux_LBC}
    \textbf{J}_L\cdot\textbf{n}=0
\end{equation}
This zero flux condition is assumed by default in COMSOL Multiphysics when a Dirichlet boundary condition is not defined (\texttt{No flux}). However, the convective term must be included in $\textbf{J}_L$ for the insulated problem to be consistent:
\begin{equation}\label{eq:Flux_ShBC}
    (-D_L\nabla C_L+\textbf{v}C_L)\cdot\textbf{n}=0
\end{equation}
See Section \ref{Diff-convect-reaction_implementation} for suitable definitions of $\textbf{v}$, which account for the role of hydrostatic stresses in driving hydrogen transport. If the stress-dependent drift term is not included, by using Eq. (\ref{Eq:Flux_LBC}) as opposed to Eq. (\ref{eq:Flux_ShBC}), very high (unrealistic) concentrations are predicted at the surface. The results obtained with the present framework are shown in Fig. \ref{fig: Kotake_softening}, for various choices of $\xi$ (or $\zeta$). The results obtained by Kotake et al.  \cite{Kotake2008TransientLoading} are also included, showing a perfect agreement. The effect of hydrogen-induced softening is accurately captured, with the lower $\xi$ values resulting in lower crack tip stresses and therefore lower $C_L$ values.

\begin{figure}[H]
  \makebox[\textwidth][c]{\includegraphics[width=0.8\textwidth]{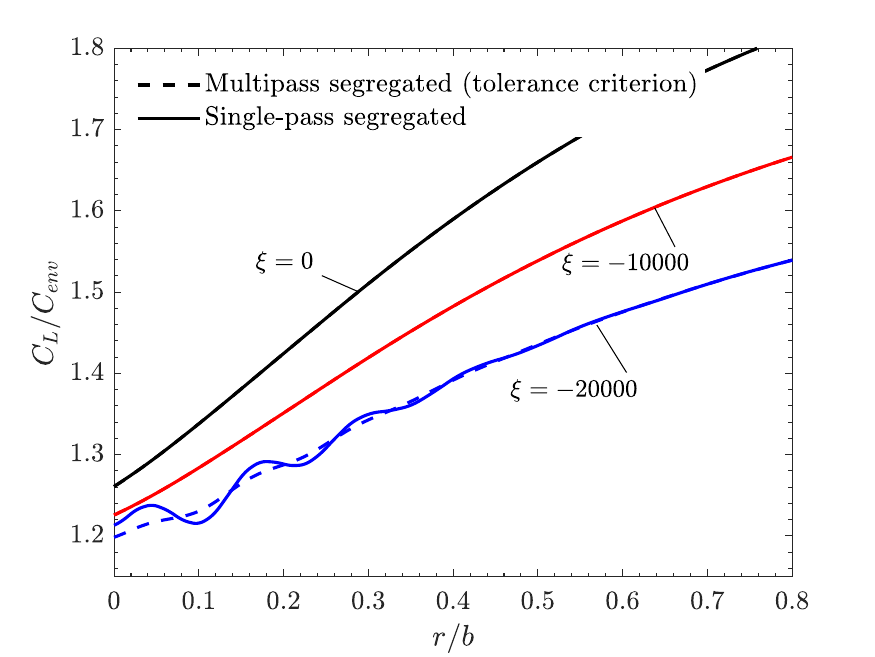}}%
  \caption{Lattice hydrogen concentration very close to the crack tip, considering different degrees of hydrogen-induced softening. A single-pass segregated (staggered) approach gives accurate results but reveals stress oscillations near the crack tip.}
  \label{fig: Kotake_tolerance_multipass}
\end{figure}

In contrast to the previous simulations, the coupling for the hydrogen-induced softening is bidirectional: stress and strain variables drive diffusion and trapping but the material behaviour is also modified by hydrogen concentration. Therefore, the sensitivity to the solution scheme is analysed. The same concentration distributions have been obtained considering a segregated or a fully coupled approach for the number of increments considered (those considered by the \texttt{Free} solver). Therefore, the subdivision of the coupled problem in a staggered scheme does not reduce the accuracy of the problem, provided that the number of solver steps is sufficiently large. However, for the scenario of strong hydrogen-induced softening ($\xi = -20000$), stress oscillations appear in deformed elements near the crack tip even with the high-order displacement discretization. This occurs for both fully coupled and segregated schemes if a single-pass step is considered, i.e. only one iteration for the complete step once each individual step subdivision converges. This is shown in Fig. \ref{fig: Kotake_tolerance_multipass}, where it is also observed that these oscillations are avoided in the segregated scheme if multiple iterations are considered until the tolerance criterion is verified, not only for each individual variable but also for the full segregated problem  (so-called multi-pass approach).  

\section{Conclusions}
\label{Sec:Concluding remarks}

We have presented a generalised framework to model hydrogen transport ahead of crack tips. The framework encompasses a very wide range of phenomena and models, bringing together all the main developments in the hydrogen embrittlement community. This generalised framework is numerically implemented in the commercial finite element package \texttt{COMSOL} and the codes are made freely available to the community. Insight is gained into the numerical challenges associated with coupled deformation-diffusion, identifying suitable stability, interpolation and solution schemes that maximise efficiency, robustness and accuracy. This work contributes to the development, dissemination and extension of models that can give valuable insights into understanding hydrogen transport and accumulation phenomena, providing a foundation for further research in the field of hydrogen embrittlement and the starting point for a hydrogen-informed phase field fracture model, as developed in Part II of the present work.

\section*{Acknowledgements}
\label{Acknowledge of funding}

\noindent The authors gratefully acknowledge funding from projects PID2021-124768OB-C21 and TED2021-130413B-I00. This work was also supported by the Regional Government of Castilla y León (Junta de Castilla y León) and by the Ministry of Science and Innovation MICIN and the European Union NextGenerationEU / PRTR through projects H2MetAmo (C17.I01.P01.S21) and MA2TEC (C17.I01). E. Mart\'{\i}nez-Pa\~neda acknowledges financial support from the EPSRC (grant EP/V009680/1), from UKRI’s Future Leaders Fellowship programme [grant MR/V024124/1], and from the UKRI Horizon Europe Guarantee programme (ERC Starting Grant \textit{ResistHfracture}, EP/Y037219/1).


\appendix
\section{Modelling Thermal Desorption Spectroscopy}
\label{App:TDS}

Trapping energies are usually characterised experimentally by means of Thermal Desorption Spectroscopy (TDS) experiments, also known as Transport Desorption Analysis (TDA). In this method, metallic samples pre-charged with hydrogen until saturation are subjected to a heating ramp $\phi$. Hydrogen desorption is measured and the observed spectra peaks are correlated to detrapping energies. While Kissinger's method \cite{Kissinger1957ReactionAnalysis}, a simplistic kinetic model that assumes infinitely fast diffusion, is frequently adopted, this approach is known to underestimate trapping energies and a more rigorous numerical approach is desirable \cite{Diaz2020InfluenceMetals, Drexler2021CriticalSpectra}. In this Appendix, we will show how the framework presented here, so far focused on hydrogen transport near a crack tip, can readily be used to model TDS experiments, where there is no coupling with mechanical loading.\\

The present framework can simulate TDS experiments using two descriptions of trapping: Orani's equilibrium or McNabb and Foster's trapping kinetics formulation. In both cases, and different to the previous isothermal case studies, the sensitivity of the diffusion coefficient to temperature has to be captured. Since temperature $T$ evolves as a function of $t$ as $T=T_0+\phi t$, the lattice diffusion coefficient can be expressed as,
\begin{equation}
    D_L = D_L^0 \exp \left[ \frac{-E_L}{R(T_0+\phi t)} \right]
\end{equation}
where $D_L^0$ is the pre-exponential diffusion coefficient, $E_L$ the activation energy for lattice diffusion and $T_0$ the initial temperature. When Oriani's equilibrium is imposed between lattice and trapped hydrogen, trapping can be modelled through the reaction term presented in Section \ref{Diff-convect-reaction_implementation}, including both $\partial C_T/\partial C_L$ and $\partial C_T/\partial T$:
\begin{equation}
        R_T = -\frac{\partial C_T}{\partial C_L}\frac{\partial C_L}{\partial t} - \frac{\partial C_T}{\partial K_T}\frac{\partial K_T}{\partial T}\frac{\partial T}{\partial t}
\end{equation}

Operating and considering that $\partial T/\partial t$ is equal to $\phi$ during TDS testing:
\begin{equation}
\label{Eq: reaction dCTdT}
         R_T = -\frac{C_T(1-\theta_T)}{C_L}\frac{\partial C_L}{\partial t} + \frac{C_T(1-\theta_T)E_B}{R(T_0+\phi t)^2}\phi
\end{equation}

And, noting that there is no hydrostatic stress, the resulting PDE is,
\begin{equation}
    \frac{\partial C_L}{\partial t}- \nabla \cdot (D_L\nabla C_L) = -\frac{C_T(1-\theta_T)}{C_L}\frac{\partial C_L}{\partial t} + \frac{C_T(1-\theta_T)E_B}{R(T_0+\phi t)^2}\phi
\end{equation}

To calculate $C_T$, and thus $\theta_T = C_T/N_T$, the dependence of $K_T$ on the current temperature is also considered, such that $K_T=\exp (E_B / (R T))$. One can also enrich Oriani's model to account for a possible faster vibration frequency of hydrogen in lattice sites (versus trapping sites). This can be captured through the ratio between pre-exponential kinetic constants $\kappa_0/\lambda_0$, which were introduced in Section \ref{eq:MFtheory}. It is generally assumed that $\kappa_0/\lambda_0=1$ and the binding energy $E_B$ and the trap density $N_T$ are the only relevant trapping parameters in the context of Oriani's model. Nevertheless, we here explore the influence of $\kappa_0$ and $\lambda_0$ and define a richer description of Oriani's model, where the equilibrium constant is given by,
\begin{equation}
\label{Eq. KT}
    K_T = \frac{\kappa_0}{\lambda_0}\exp\left[ \frac{E_B}{R(T_0+\phi t)} \right]
\end{equation}

If equilibrium cannot be assumed, McNabb and Foster's formulation, Eq. (\ref{eq. MCNabb-Foster}), is used but considering the temperature dependence of kinetic constants:
\begin{equation}
    \kappa = \kappa_0 \exp\left[ \frac{-E_t}{R(T_0+\phi t)} \right]
\end{equation}
\begin{equation}
    \lambda = \lambda_0 \exp\left[ \frac{-E_d}{R(T_0+\phi t)} \right]
\end{equation}
where $E_t$ and $E_d$ represent trapping and detrapping energies, respectively. Their relationship with the binding energy that determines equilibrium is $E_B = E_d - E_t$.\\

To simulate TDS experiments, a 1D geometry is considered, where hydrogen desorption from a slab of thickness $d$ is simulated. Only half of the slab ($d/2$) is simulated with a zero flux as a symmetry condition and a zero concentration is imposed in the outer surface node, i.e. $C_{env}=0$, where desorption occurs. The slab is discretised with 1000 elements, with the mesh being finer near the outer node, where higher gradients are expected. Quadratic discretization is chosen for $C_L$, and also for $C_T$ when kinetic trapping is considered.\\

First, we validate the TDS predictions of our dffusion-trapping framework against the results by Legrand et al. \cite{Legrand2015TowardsModeling}. While their work is based on a kinetic trapping description, we provide results for both Oriani and McNabb and Foster models. The parameters employed are listed in Table \ref{Tab:TDSparameters}. Traps are considered to be initially in equilibrium and the initial occupancy, $\theta_T^0$, is determined from $C_L^0$ and $K_T$. However, due to the extremely low initial temperature ($T_0=$ 10 K) assumed, the value of $K_T$ at the beginning of the analysis is very high and consequently $\theta_T^0 \approx 1$.

\begin{table}[H]
\centering
\caption{TDS model parameters, following Ref. \cite{Legrand2015TowardsModeling}.}
\label{Tab:TDSparameters}
   {\tabulinesep=1.2mm
   \makebox[\textwidth][c]{\begin{tabu} {cccccccccccc}
       \hline
 $D_L^0$ & $N_T$ & $E_L ; E_t$ & $E_d$ & $\kappa_0$ & $\lambda_0$ \\ \hline
 \num{2.74e-6} & 2.0 & 0.2 & 0.6 & $10^{13}$  & $10^8$  \\
 (m$^2$/s) & (mol/m$^{3}$) & (eV) & (eV) & (s$^{-1}$)   &  (s$^{-1}$) \\\hline
   \end{tabu}}}
   {\tabulinesep=1.2mm
   \makebox[\textwidth][c]{\begin{tabu} {cccccccccccc}

 $d$   &  $T_0$ & $\phi$ & $C_L^0$ & $N_L$ \\ \hline
 4 & 10 & 50 &  1.0 &   \num{2.1e5} \\
 (mm) & (K) & (K/min) & (mol/m$^{3}$) & (mol/m$^{3}$)\\\hline
   \end{tabu}}}
\end{table}

\begin{figure}[H]
  \makebox[\textwidth][c]{\includegraphics[width=0.8\textwidth]{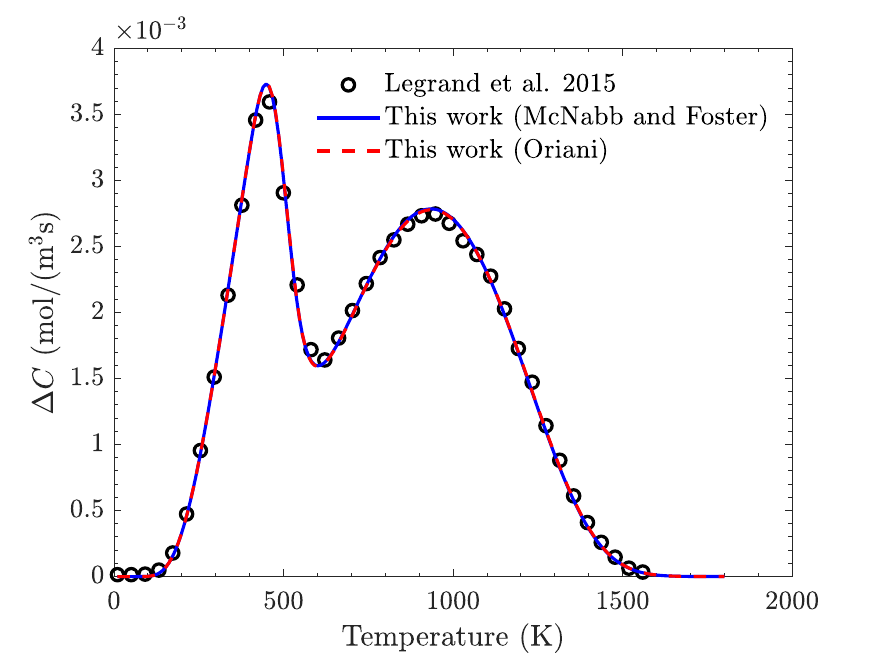}}%
  \caption{Hydrogen desorption measured as the variation in total hydrogen concentration during a simulated TDS test. The evolution predicted by two different formulations, McNabb and Foster's kinetic trapping or Orani's equilibrium, is compared to results from Ref. \cite{Legrand2015TowardsModeling}.}
  \label{fig: Validation Legrand}
\end{figure}

As shown in Figure \ref{fig: Validation Legrand}, results perfectly agree with those from the original benchmark for both modelling assumptions, equilibrium or kinetic trapping. 

\begin{figure}[H]
  \makebox[\textwidth][c]{\includegraphics[width=0.8\textwidth]{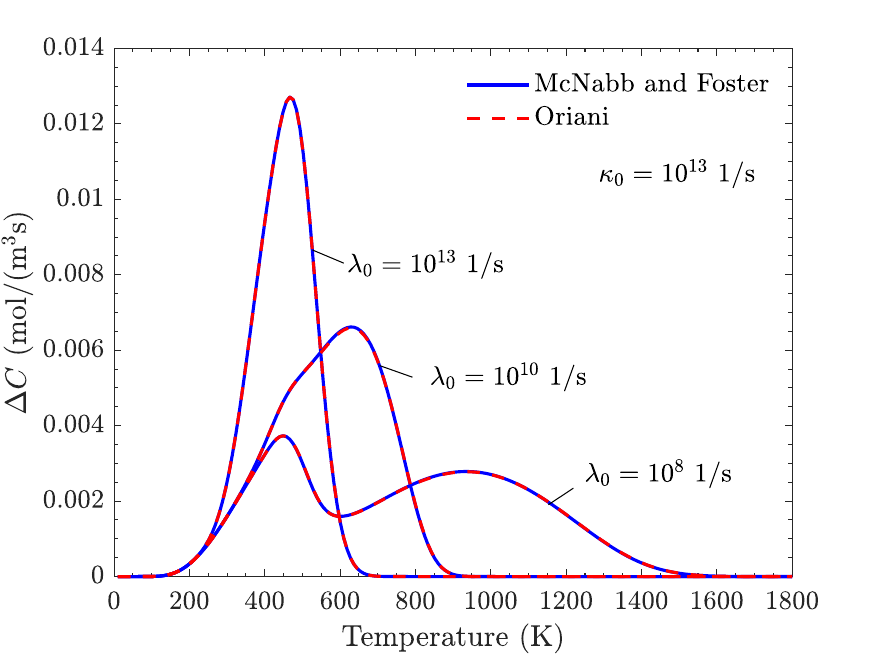}}%
  \caption{Influence of detrapping frequencies ($\lambda_0$) on the shift ot TDS spectra. Results assuming Oriani's equilibrium  are obtained with a consistent reaction rate, i.e. including $\partial C_T / \partial T$.}
  \label{fig: TDS lambda influence}
\end{figure}

Strong trapping in that benchmark case is not only a result of the $E_d$ value, 0.6 eV or 57.9 kJ/mol, but also a consequence of the assumed pre-exponential kinetic constants. A trapping frequency $\kappa_0$ was assumed to be much higher than the detrapping frequency $\lambda_0$ by Legrand et al. \cite{Legrand2015TowardsModeling} to obtain realistic desorption times. If the trapping frequency $\kappa_0$ is reasonably fixed as the Debye frequency \cite{Krom2000HydrogenSteel}, i.e. $10^{13}$ s$^{-1}$, the influence of $\lambda_0$ values is shown in Figure \ref{fig: TDS lambda influence}. Higher release frequencies produce faster detrapping and earlier peaks that can be merged with the peak corresponding to lattice desorption. The same results are obtained with Oriani's equilibrium or with McNabb and Foster's kinetic formulation (Figure \ref{fig: TDS lambda influence}). It must be noted that if equilibrium is assumed, the $C_T$ variation with temperature, i.e. last term in Eq. (\ref{Eq: reaction dCTdT}), must be explicitly included. Frequency values do not directly influence the transport equation if Oriani's equilibrium is chosen, but the ration $\kappa_0/\lambda_0$ influences $K_T$, according to Eq. (\ref{Eq. KT}). However, a different vibration frequency in lattice or trapping sites, i.e.  $\kappa_0/\lambda_0\neq 1$,  would need further justification.\\

\begin{figure}[H]
  \makebox[\textwidth][c]{\includegraphics[width=0.8\textwidth]{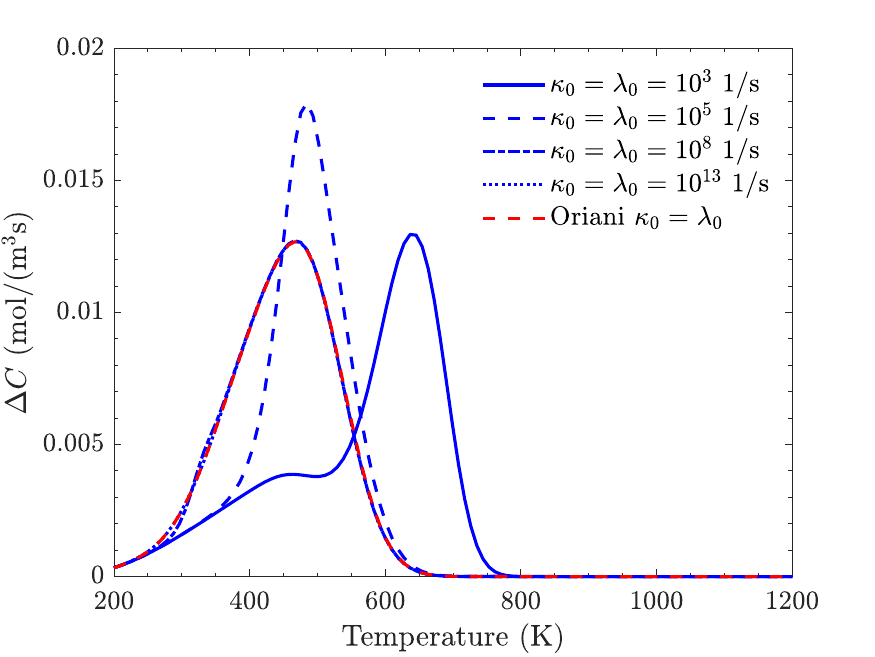}}%
  \caption{Influence of vibration frequencies on TDS peaks when $\kappa_0$ and $\lambda_0$ are equal. Different frequency values are simulated considering McNabb and Foster's formulation while the model based on Oriani's equilibrium only depends on the $\kappa_0/\lambda_0$ ratio.}
  \label{fig: TDS influence frequencies}
\end{figure}

The validity of Oriani's equilibrium is verified for high frequencies, but it is here observed, as in Ref. \cite{Diaz2020InfluenceMetals}, that McNabb and Foster's formulation is sensitive to the choice of $\kappa_0$ and $\lambda_0$ values and not only of their ratio. Fixing $\kappa_0 = \lambda_0$, low frequencies delay the attainment of equilibrium and therefore TDS peaks are shifted to higher temperatures - this is shown in Fig. \ref{fig: TDS influence frequencies}. In contrast, for a vibration frequency higher than $10^8$ s$^{-1}$, all results converge to the spectra predicted by Oriani's equilibrium, where only the ratio $\kappa_0/\lambda_0$ influences the process.

\begin{figure}[H]
    \centering
    \begin{subfigure}[h]{0.49\textwidth}
        \centering
        \includegraphics[width=\textwidth]{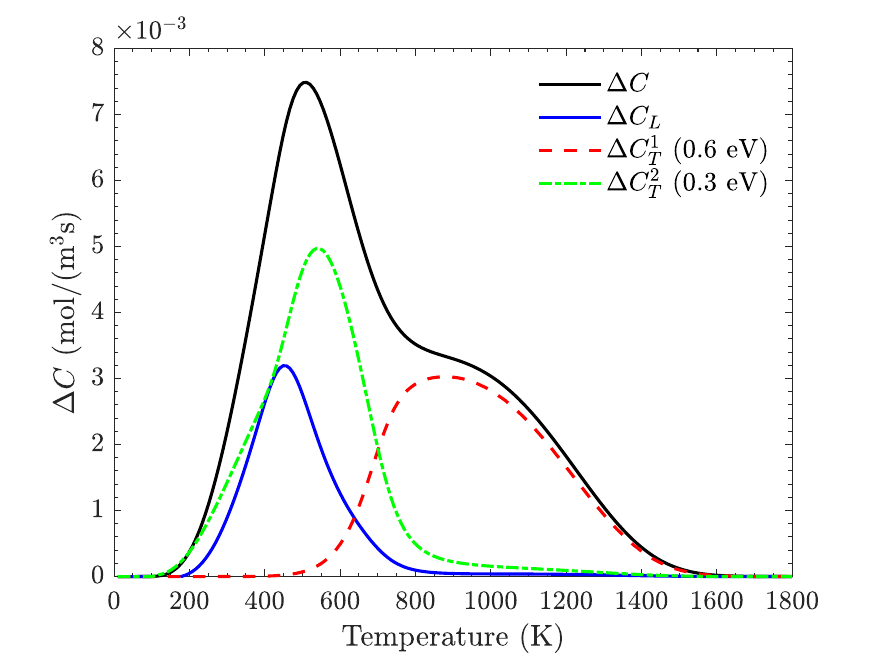}
        \subcaption{}
        \label{}
    \end{subfigure}
    \begin{subfigure}[h]{0.49\textwidth}
        \centering
        \includegraphics[width=\textwidth]{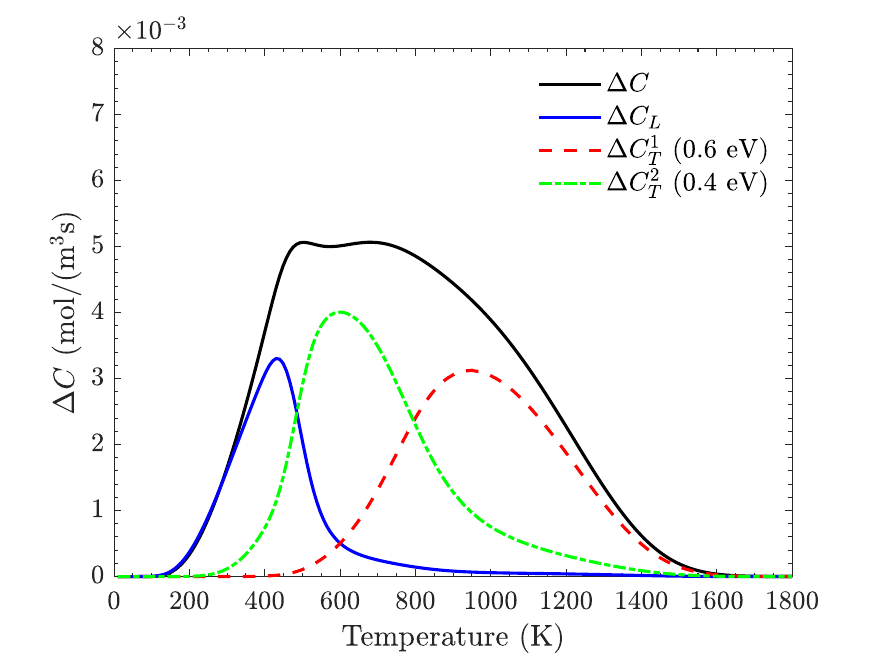}
        \subcaption{}
        \label{}
    \end{subfigure}
    \begin{subfigure}[h]{0.49\textwidth}
        \centering
        \includegraphics[width=\textwidth]{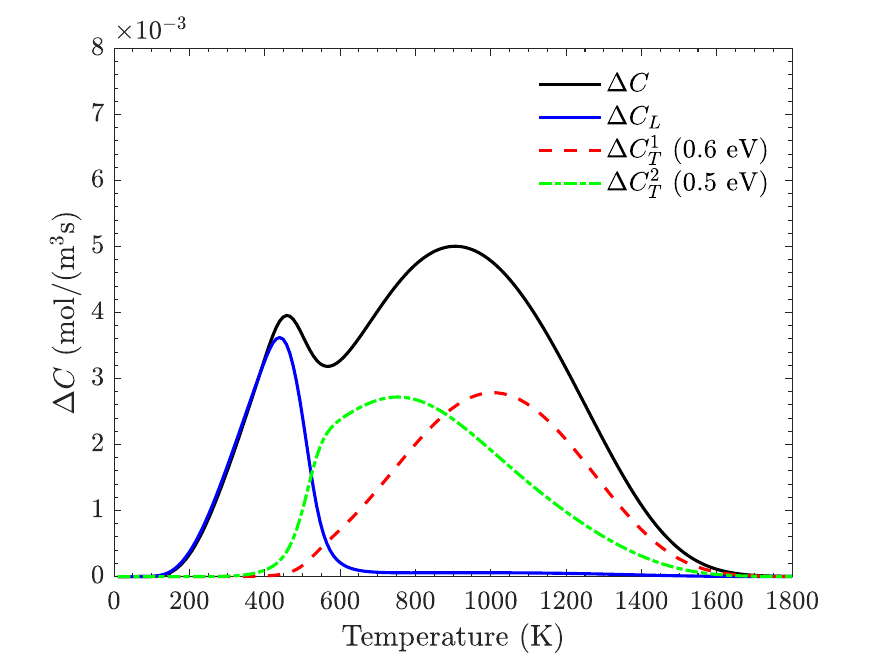}
        \subcaption{}
        \label{}
    \end{subfigure}
    
    \caption{Evolution of total, lattice and trapped hydrogen during TDS considering two trapping sites. The first trapping site has a detrapping energy of $E_d^1$ = 0.6 eV whereas the second trap is assessed with (a) $E_d^2$ = 0.3 eV, (b) $E_d^2$ = 0.4 eV, and (c) $E_d^2$ = 0.5 eV.}
    \label{fig: TDS 2 traps}
\end{figure}

Finally, the case proposed by Legrand et al. \cite{Legrand2015TowardsModeling} is extended to include a second trap site, with the corresponding extra PDE to model kinetic trapping. Therefore, two defects are simulated: trap 1 with the previously used values, $N_T^1$ = 2 mol/m$^3$ and $E_d^1$ = 0.6 eV, and a new trapping site with $N_T^2$ = 2 mol/m$^3$ and $E_d^2$ = 0.3, 0.4 or 0.5 eV. The weaker second trap modifies the trapping, detrapping and desorption phenomena as can be observed in Figure \ref{fig: TDS 2 traps} for the different $E_d^2$ simulated values.

%\bibliographystyle{elsarticle-num}

%\bibliography{library}
\end{document}